\documentclass[apj]{emulateapj}

\usepackage{graphicx}
\usepackage{latexsym}
\usepackage{amsmath,amssymb}
\usepackage{natbib}
\usepackage{dcolumn}
\usepackage{bm}
\usepackage{calrsfs}
\usepackage[hypertex]{hyperref}
\usepackage{ulem}
\usepackage[usenames]{color}

\newcommand{\myemail}{bmv@mpi-hd.mpg.de}

\def\be{\begin{equation}}
\def\ee{\end{equation}}
\def\ba{\begin{eqnarray}}
\def\ea{\end{eqnarray}}
\def\d{\delta}

\def\k{\varkappa}

\def\b{\boldsymbol}

\def\sun{\odot}
\def\pks{\object{PKS~2155$-$304}}
\def\mkn501{\object{Mkn~501}}
\def\pkstitle{PKS~2155$-$304}
\newcommand{\na}{New Astr.}

\shorttitle{Rapid variability due to Jet-Star interaction}
\shortauthors{Barkov et al.}

\begin{document}


\title{Rapid TeV variability in Blazars as result of Jet-Star Interaction}


\author{M.V.~Barkov\altaffilmark{1,2}, F.A.~Aharonian\altaffilmark{3,1}, S.V.~Bogovalov\altaffilmark{4}, 
S.R.~Kelner\altaffilmark{1,4}, and D.~Khangulyan\altaffilmark{5}}
\affil{${}^1$Max-Planck-Institut f\"ur Kernphysik,
Saupfercheckweg 1, D-6917 Heidelberg, Germany; \myemail}
\affil{${}^2$Space Research Institute RAS, 84/32 Profsoyuznaya Street, Moscow, 117997, Russia}
\affil{${}^3$Dublin Institute for Advanced Studies, 31 Fitzwilliam Place,
Dublin 2, Ireland}
\affil{${}^4$National Research Nuclear University (MEPHI), Kashirskoe shosse
31, 115409 Moscow, Russia}
\affil{${}^5$Institute of Space and Astronautical Science/JAXA, 
3-1-1 Yoshinodai, Chuo-ku, Sagamihara, Kanagawa 252-5210, Japan}




\begin{abstract}
  We propose a new model for the description of ultra-short flares
  from TeV blazars by compact magnetized condensations (blobs),
  produced when red giant stars cross the jet close to the central
  black hole.  Our study includes a simple dynamical model for
  the evolution of the envelope lost by the star in the jet, and its
  high energy nonthermal emission through different leptonic and
  hadronic radiation mechanisms.  We show that the fragmented envelope
  of the star can be accelerated to Lorentz factors up to 100 and
  radiate effectively the available energy in gamma-rays predominantly
  through proton synchrotron radiation or external inverse Compton
  scattering of electrons.  The model can readily explain the
  minute-scale TeV flares on top of longer (typical time-scales of
  days) gamma-ray variability as observed from the blazar \pks.  
  In the framework of the proposed scenario, the key
  parameters of the source are robustly constrained. In the case of
  proton synchrotron origin of the emission a mass of the central
  black hole of $M_{\rm BH}\approx 10^8 M_{\odot}$, a total jet power
  of $L_{\rm j} \approx 2\times 10^{47} \, \rm erg\,s^{-1}$ and a
  Doppler factor, of the gamma-ray emitting blobs, of $\delta\geq 40$
  are required. Whilst for the external inverse Compton model, parameters of
  $M_{\rm BH}\approx 10^8 M_{\odot}$, $L_{\rm j} \approx 10^{46} \,
  \rm erg\,s^{-1}$ and the $\delta\geq 150$ are required.
\end{abstract}





\section{Introduction}

The flux variability of very high energy (VHE) gamma-rays on minute
timescales detected from the BL Lac object \pks{} \citep{ah07pks} 
and Mkr 501 \citep{mkr501_magic}
challenges the standard scenarios suggested for the explanation of the nonthermal
properties of TeV blazars \citep{bfr08,gub09}. The extremely short duration
of the flares impose severe constraints on the size of the gamma-ray producing region, of
\be
l'\le c \tau'\simeq 3\times10^{13}\tau'_{3}\,\rm cm,
\label{eq_size_prop}
\ee
where $l'$ and $\tau'_{3}=\tau'/10^3 $~s are the proper
production size and the variability timescale in the frame of the jet
respectively and $c$ is the light speed. The proper variability time-scale, $\tau'$,
is connected to the variability in the observer frame, $\tau$,
by the relation
\be
\tau= \frac{\tau'}{ \delta} \, ,
\label{eq_var_tran}
\ee
 where $\delta$ is the Doppler factor of the moving source (the blob):
\begin{equation}
\delta=\frac{1}{\Gamma\left(1-\beta\cos(\alpha)\right)} \ .
\label{db}
\end{equation}
Here the bulk Lorentz factor, $\Gamma$, accounts for the
relativistic transformation of time, and $\left(1-\beta\cos(\alpha)\right)$
is responsible for the kinematic shrinking
of the duration of the radiation and $\beta=v/c$.

Relativistic jets ejected from the central engines are common
phenomena for different types of active galactic nuclei (AGN). In
particular, apparent superluminal speeds ${\beta_{\rm
 app}=\beta\sin{\alpha}/(1-\beta\cos{\alpha})}$ (in units of the
speed of light~$c$) as high as $\sim 40$ have been detected for radio
components on (projected) scales of $\sim 1-10\;$pc
\citep[see e.g.][]{J01,J05,mjl10} in blazars - AGN
with jets directed to the observer.  This implies very large
Lorenz factors of bulk motion given that $\Gamma \geq \beta_{\rm
 app}$.

Because of the large bulk Lorenz factors of TeV blazars,
the condition Equation~(\ref{eq_var_tran}) allows significant relaxation of the requirement supplied by
Equation~(\ref{eq_size_prop}). In particular,
\citet{lb08} argued that if a perturbation is produced by the central
engine, its size should exceed the gravitational radius of the black
hole in the observer frame. Consequently, the proper size of the
production region is expected to be larger than $\Gamma r_ {\rm g}$,
where $r_{\rm g}=GM_{\rm BH}/c^2 \approx 1.5 \times 10^{13} M_{\rm BH,8}$ cm is the
gravitational radius of a black hole (BH) of mass
$M_{\rm BH,8}=M_{\rm BH}/10^8 M_\odot$. In this case, the variability
time-scale $\tau_2=\tau/10^2$s imposes a strict upper limit on the
gravitational radius: 
\be 
r_{\rm g}<3\times10^{12} \tau_2 \frac{\delta}{ \Gamma}\, \rm cm \ .
\label{rg}
\ee
Thus, the detection of gamma-ray flux variability $\tau \sim
200$~s constrains the BH mass, to be $M_{\rm BH,8} <1$. In reality, since the
main energy release occurs in the inner parts of the accretion disk of
radius $r \sim 10 r_{\rm g}$, the mass of the black hole should be
close to $10^7 M_\odot$. Generally, in the case of an extreme Kerr
black hole, the energy release takes place at the gravitational
radius. However, even in this case one needs an entire rotation period for
an effective energy release, i.e. the characteristic time cannot be
much shorter than $2 \pi r_{\rm g}/c$. This implies that even in this
case the upper bound of the BH mass of $10^7 M_\odot$ cannot be
significantly relaxed.

This conclusion is true, in particular, for the model of internal
shocks. Note, however, that it is based on the assumption
that the perturbation (the reason of the flare) originates in the central engine. 
Therefore it cannot be unconditionally
extended to other possible scenarios as is claimed by
\citet{dfkb09}. Indeed, Equation~(\ref{rg}) is not valid if
perturbations are produced by an external source, e.g. by plasma
condensations (often called "blobs") which do not have a direct link to
the central black hole. Such blobs can be produced, in particular, by
interactions of stars with the base of the jet, as proposed by
\citet{bab10} to explain the TeV flares of M~87 on scales
of days \citep[see also][]{abr10}, where the interaction of gas
clouds from broad line regions are discussed). The jet power of M~87 is
relatively modest, $L_{\rm j} \simeq 10^{44}\,\rm erg\,s^{-1}$.  The
results of \citet{bab10} show that, while this power is
sufficient to blow-up the envelope (atmosphere) of the star which
initially has been pulled out by the tidal force of the BH, such a jet
appears to be not sufficiently powerful for acceleration of the gas
cloud to relativistic velocities. Actually this works in a positive
direction for M~87, given the large aspect angle of the
jet. Otherwise, the gamma-ray flux could not be observed because of
the Doppler de-boosting effect. On the other hand, the suggested
mechanism of formation of hadronic blobs in the jets cannot apply to
powerful gamma-ray blazars unless the blobs are accelerated to Lorentz
factors $\Gamma \geq 10$.  Remarkably, this can be realized in a quite
natural way in powerful jets with $L_{\rm j} \geq 10^{46}\,\rm
erg\,s^{-1}$.  Interestingly, such powerful jets can ablate the star
atmosphere without help of tidal forces (the interesting
implications of this effect are discussed below).  Moreover, the
powerful jets can drag and disrupt the star's envelope into an 
ensemble of blobs moving with large Lorentz factors which, 
from the point of view of explanation of very short time variability
is a quite comfortable situation. 


Another important aspect of the short time variability is
related to the efficiency of acceleration and radiation mechanisms.
Currently the most conventional approach for
modeling of VHE emission production in active galaxies is based on
the inverse Compton (IC) scattering of relativistic electrons,
the soft target photon field being either the synchrotron radiation of
same electrons (the so-called Synchrotron-self Compton (SSC) model),
or provided by external sources (EIC model). The apparent advantage of IC
models is the combination of two factors: (1) the acceleration of electrons to
relatively modest energies ($\leq 1$~TeV) can be effectively realized
within different acceleration scenarios; and (2) these electrons radiate readily
in interactions of ambient radiation and magnetic fields. Nevertheless,
while the IC models allow rather satisfactory explanations of
the energy spectra and variability patterns of many blazars in general,
the parameters used to fit some specific objects appear incompatible with the parameters
defined from  observations. Moreover, the observed short
variability time-scale demands conditions which appear to be quite uncomfortable,
in terms of the strength of magnetic field and related
consequences concerning the
strong deviation from equipartition between the energy density of relativistic electrons and
magnetic fields. The requirement of weak, (generally less than 1 G) magnetic fields
is one of the key postulates of the IC paradigm of gamma-ray production in blazars.
Moreover, in the case of some objects with unusually hard source spectra (after correction
for intergalactic absorption), such as 1ES~0229+200, the magnetic field is required
to be as small as 1mG \citep{tgg09}. The magnetic field in the blazar
jets can be reduced to such small values only at very large distances from the
central engine, namely $\gtrsim 10^{18} \ \rm cm$. Although this idea has some observational
support related to the transparency of blobs in the radio band \citep{mjd08}, it is likely that
regions of highly variable gamma-ray emission are located at smaller distances from the
central engine \citep{tgb10}. In particular, the EIC models require the location of any gamma-ray
emitter to be located closer to the BH into the so-called Broad Line Regions (BLR), 
i.e typically at distances
$R \sim 10^{17} - 10^{18}$~cm. This implies that any IC model can be realized only if
one finds a way dramatically reduce the magnetic field in the jet.
Although this cannot be excluded (e.g. because of reconnection of the B-field
\citep{kbl07,gub09,mu10} or due to the effective bulk acceleration of plasma
\citep{kvk10,tnm10}), strong magnetic fields exceeding 1~G remain a more
favored option, as long as we deal with strong jets on sub-parsec scales.
In this regard, the models which invoke high energy protons for the production
of gamma-rays passes  certain advantages despite a quite popular view that they are
not effective emitters \citep[see e.g.][]{s10psyn}. Actually this is true only for
proton-proton and proton-photon interactions. What concerns the synchrotron
radiation of protons, with a key assumption on the acceleration of particles
with a rate close to $t^{-1} \sim ceB/E$, where $E$ is the proton energy,
coupled with a strong magnetic field between 10 to 100 Gauss, and a large Doppler
factor, $\delta \geq 10$, it is that can provide relevant acceleration and
radiation timescales, as well as explain the extension of gamma-ray
spectra to TeV energies \cite[see e.g.][]{ah00}.

\section{Blobs in Relativistic Jets}
\subsection{AGN Jet -- Red Giant interaction}
\begin{figure}
\plotone{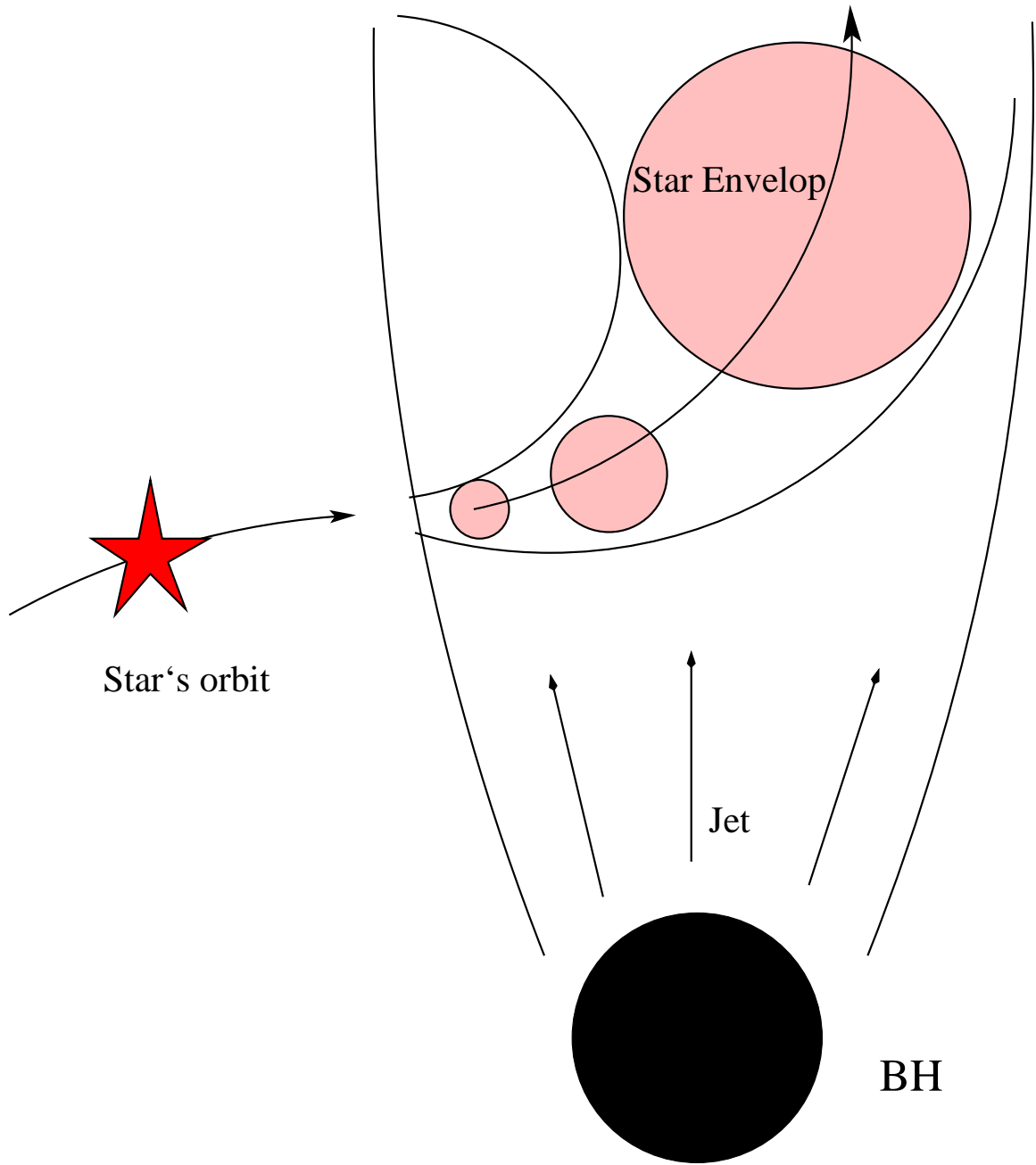}
\caption{Schematic illustration of the scenario. If a star enters a
  powerful AGN jet, the outer layers of the star atmosphere are to be
  ablated.  Because of the interaction with the jet the ablated cloud
  expands and gets involved into the jet bulk motion.  }
\label{sk1}
\end{figure}

Below we discuss the distinct features of the interaction of red giants with
AGN jets in the specific case of powerful blazars, see sketch of the scenario in
Figure~\ref{sk1}.
Originally, the scenario of AGN~jet -- red~giant (RG) interaction
has been suggested by \citet{bab10} for the explanation of
VHE observations of M87 - a nonblazar type nearby AGN with
a large jet viewing angle of $\geq 20^\circ$ and a modest jet power,
of $L_{\rm jet} \simeq 10^{44} \,\rm erg\,s^{-1}$. It was demonstrated that if  
disturbed by tidal forces RG penetrates the jet, the ram pressure of the
jet in M87 would be sufficient to remove the outer
layer of the RG. This leads to the formation of a dense cloud within the jet which,
in combination with effective particle acceleration,
can trigger gamma-ray production through proton-proton interactions
which passes secure of the required spectral and temporal properties. This model allows
detectable gamma-ray fluxes because of the proximity of the source
and the non-relativistic speed of the blobs (otherwise the radiation from M87 would be
de-boosted given the large aspect angle of the jet).

The AGN jet -- red giant interaction (JRGI) has very specific and important
features in the case of powerful blazars, where the ram
pressure can be as high as
\begin{equation}
P_{\rm ram}\simeq10^3 L_{\rm j,46}z_{17}^{-2}\theta_{-1}^{-2}\, \rm dyn\, cm^{-2} \, .
\label{eq:ram}
\end{equation}
Here $L_{\rm j,46}=L_{\rm j}/10^{46}\rm erg\, s^{-1}$,
$z_{17}=z/10^{17}\rm cm$ and $\theta_{-1}=\theta/0.1$ are the jet
power, the distance from the BH, and the jet opening angle, respectively.
Such a high ram pressure can blow-up the outer layers of the stellar atmosphere,
even from a non-disturbed RG. The mass of the removed layer can be roughly
estimated from the balance of the jet ram force to the gravitational force,
$P_{\rm ram}\pi R_*^2\simeq G\Delta M M_*/R_*^2$. Here $M_*$ and $R_*$
are the RG mass and radius, respectively. This gives the
mass of the cloud stripped from the star by the jet:
\begin{equation}
\Delta M=\frac{\pi P_{\rm ram} R_{*}^4}{G M_{*}}
\approx 6\times 10^{28} L_{\rm j,46}z_{17}^{-2}\theta_{-1}^{-2}R_{*,2}^4 M_{*,0}^{-1}\, \rm g,
\label{dms}
\end{equation}
where $R_{*,2}=R_{*}/10^2 R_{\odot}$ and $M_{*,0}=M_{*}/M_{\odot}$ are
the RG parameters expressed through the solar mass ($M_{\odot}$) and radius
($R_{\odot}$). This estimate illustrates the feasibility of formation of
a cloud due to the JRGI process. The next important issue is the
acceleration of the cloud in the jet. In what follows, we
assume that initially the ejected matter from the RG envelope
forms a (quasi) spherical cloud of radius comparable to the RG size,
$r_{\rm c}\simeq R_*$ .

To be accelerated, the matter residing in the jet first should be
heated, thus the cloud expands enhancing the interaction between the
cloud and the jet. A significant expansion occurs on a time-scale of
$t_{\rm exp}=A t_{\rm cc}$, where $A$ is a constant of order of 5
\citep[see e.g.][]{gmr00,nmk06,phf10}, and $t_{\rm cc}=2r_{\rm c}/c_{\rm s}$ is
the sound crossing time. Here $c_{\rm s}$ is the sound speed in the shocked
cloud, which can be estimated as $c^2_{\rm s}\approx \gamma_{\rm g}
P_{\rm ram}/\rho_{\rm c}\approx \gamma_{\rm g} P_{\rm ram} 4 \pi
r_{\rm c}^3/( 3 M_{\rm c} )$ ($\gamma_{\rm g}=4/3$ is the plasma
adiabatic coefficient). This gives the following estimate for the expansion time:
\begin{equation}
t_{\rm exp}=A t_{\rm cc}=5\times10^5 R_{*,2}^{-1/2} M_{\rm c, 26}^{1/2} L_{\rm j,46}^{-1/2}z_{17}\theta_{-1}\,\rm s,
\label{eq:expansion}
\end{equation}
where $M_{\rm c, 26} = M_{\rm c}/10^{26}\rm g$ is the normalized
mass of the cloud.
The orbital velocity of a star around the black hole,
\begin{equation}
V_{\rm orb}\approx \sqrt{\frac{G M_{\rm BH}}{z}}
\approx 4 \times 10^8 M_{\rm BH,8}^{1/2} z_{17}^{-1/2} \mbox{cm s}^{-1}\,,
\label{sv}
\end{equation}
results in a crossing time of the jet of
\begin{equation}
t_{\rm jc}\approx \frac{z^{3/2}\theta}{\sqrt{GM_{\rm BH}}}=3\times10^7z_{17}^{3/2}\theta_{-1}M_{\rm BH,8}^{-1/2}\rm s.
\label{jct}
\end{equation}
The condition $t_{\rm exp}<t_{\rm jc}$ defines an upper limit
on the mass of the cloud, which can expand in the jet:
\begin{equation}
M_{\rm c,sc}\lesssim 3\times10^{29}L_{\rm j,46} R_{*,2} z_{17} M_{BH,8}^{-1}\,\rm g.
\label{mcsc}
\end{equation}
If this constraint is fulfilled, then at the stage of hydrodynamical
expansion, the cloud can increase its linear size by a few orders of
magnitude \cite[see e.g.][]{phf10} whilst remaining in the jet.

The cloud will be trapped by the jet if it would be sufficiently accelerated
along the jet axis. Thus, the confinement condition has the following form: $v_z =
a_z t_{\rm jc} > V_{\rm orb}/\theta$, where $a_z=P_{\rm ram} \pi
r_{\rm c}^2/ M_{\rm c}$. Using Equations~(\ref{eq:ram}) and (\ref{jct}) one
obtains the cloud capture condition:
\begin{equation}
M_{\rm c,jc}\lesssim 3\times 10^{31} L_{\rm j,46} r_{\rm c,15}^2 M_{\rm BH,8}^{-1}\, \mbox{ g ,}
\label{mcjc}
\end{equation}
where $r_{\rm c,15}=r_{\rm c}/10^{15}\rm cm$ is the size of the cloud
after the hydrodynamical expansion. Importantly, Equations~(\ref{mcsc}) and
(\ref{mcjc}) provide upper limits, which exceed significantly the
expected mass of the blown-up layer given by Equation~(\ref{dms}). Thus, even if the
ablated stellar matter  remains as one
blob, this cloud will  still be trapped in the jet and accelerated up
to sub-relativistic velocities (i.e. $\sim0.1c$). In fact,
independent of the initial conditions, during the acceleration phase
the cloud is expected to be crushed into hundreds of small blobs
\citep{phf10}. This relaxes significantly the above obtained 
conditions, since the ensemble of blobs can be more easily picked up by the jet flow.

\subsection{Relativistic Stage}\label{acac}

At the relativistic stage, the dynamics of the cloud is described by the
following equation:
\begin{equation}
\frac{d\Gamma_{\rm c}}{dt}=\left(\frac{1}{\Gamma_{\rm c}^2}-\frac{\Gamma_{\rm c}^2}{\Gamma^4_{\rm j}}\right)\frac
{L_{\rm j} r_{\rm c}^2}{4\omega^2 c^2 M_{\rm c}}\,\rm,
\label{dgdt}
\end{equation}
where $\Gamma_{\rm c}$ is the Lorentz factor of the cloud,
$\omega\approx z\theta$ is cylindrical radius of the jet (see
Appendix~\ref{der12} for the derivation of Equation~(\ref{dgdt})).  Let us
introduce the following notations $ g\equiv \Gamma_{\rm c}/\Gamma_{\rm j}$ and
$y\equiv z/z_0$ (and $dy\approx c dt/z_0$). Here $z_0$ is the distance
from the BH to the point where the RG penetrates the  jet, i.e., we adopt the
initial condition $g\ll1$ at $z=z_0$. Several simplifications, in
particular, assuming that $\Gamma_{\rm j}=\rm const$ on the cloud
acceleration time scale, and $dt= dy z_0/c$, allow presentation of
Equation~(\ref{dgdt}) in the form:
\begin{equation}
\frac{d g}{dy} = \left( \frac{1}{g^2}-g^2 \right) \frac{D}{y^2}\,,
\label{dgdtb}
\end{equation}
where
\begin{equation}
D \equiv \frac{L_{\rm j} r_{\rm c}^2}{4\theta^2\Gamma_{\rm j}^3z_0 c^3 M_{\rm c}}\,.
\label{DD}
\end{equation}
Equation~(\ref{dgdtb}) allows an analytical solution, which
defines $y$ as a function of $g$ and $D$.

\begin{figure}
\plottwo{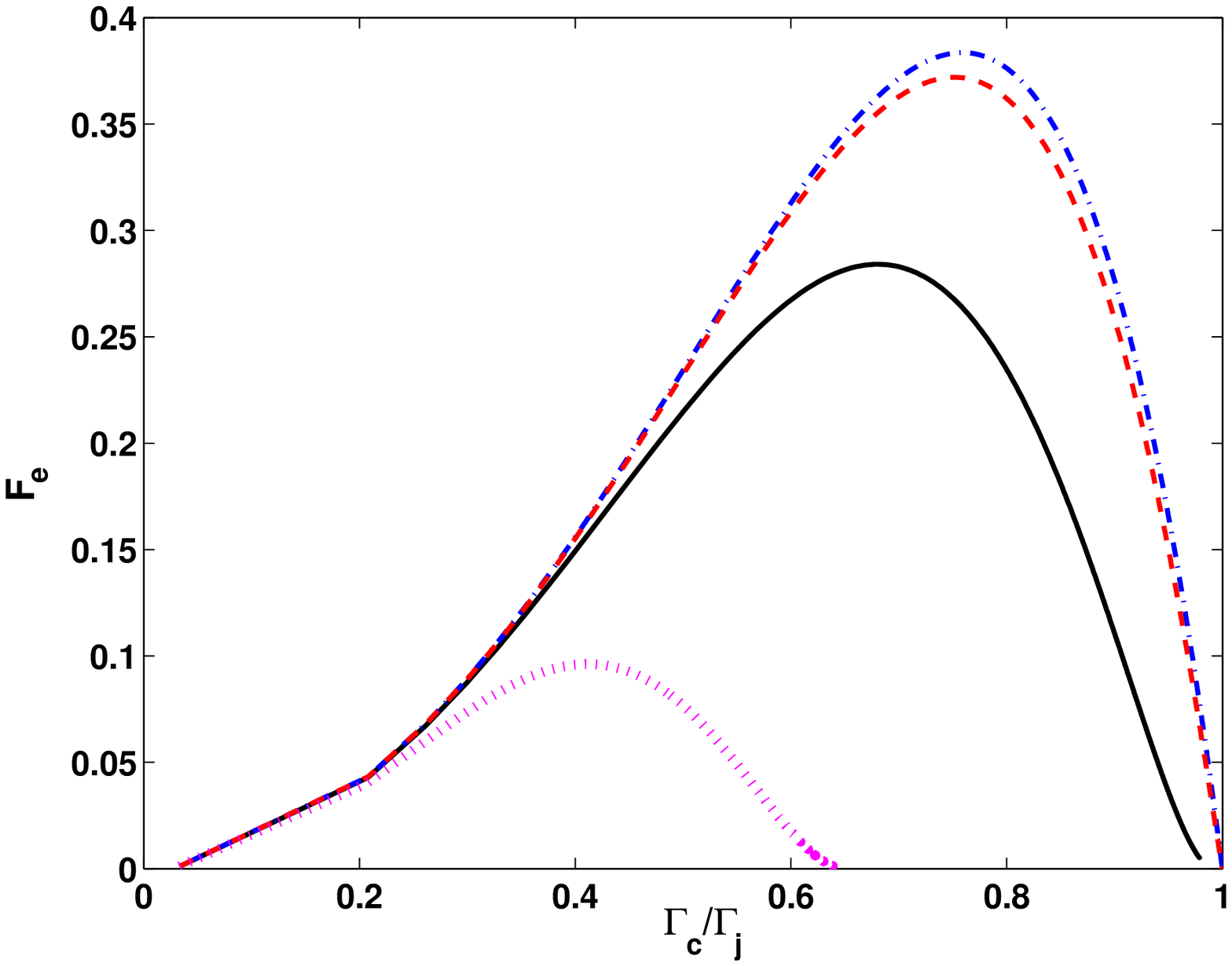}{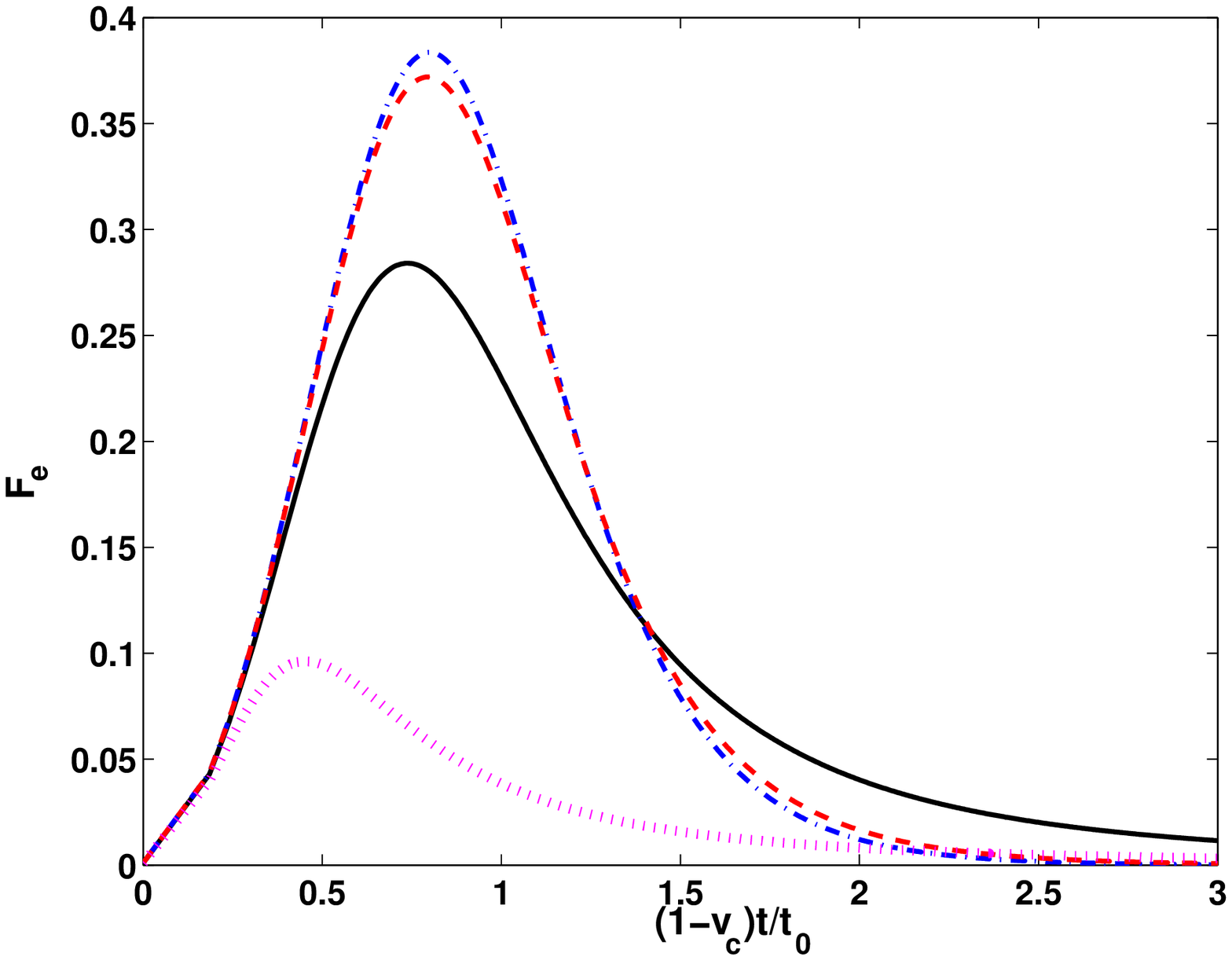}
\caption{Solutions of Equation~(\ref{dgdtb}) shown as
$F_e$ vs. the relative Lorentz factor of the cloud (left panel) and as
$F_e$ vs. the observation time (in units $t_0=z_0/2D\Gamma_{\rm j}^2c$) (right panel). The Lorentz factor of the jet
is assumed to be $\Gamma_{\rm j}=30$. The following values of the D-parameter were used: $D=100$ (dot-dashed lines),
$D=10$ (dashed line), $D=1$ (solid line) and $D=0.1$ (dotted line).}
\label{cac}
\end{figure}

In this paper we do not specify the origin of relativistic particles
in the cloud. However, assuming that particle acceleration is a result
of a strong interaction between the cloud and the jet, we may conclude
that high Lorentz factors do not support the production of non-thermal
radiation. Indeed, as it follows from Equation~(\ref{dgdtb}), the cloud-jet
interaction intensity decreases with cloud acceleration.  The apparent
intensity of the non-thermal phenomena can be roughly described by the
luminosity correction function $F_{\rm e}$, which accounts both for the
Doppler boosting, i.e. $F_{\rm e}\propto \Gamma_{\rm c}^4$, and for the
interaction intensity, i.e. $F_{\rm e}\propto
\left(\frac{1}{\Gamma_{\rm c}^2}- \frac{\Gamma_{\rm
      c}^2}{\Gamma^4_{\rm j}}\right)/z^2$. For the sake of clarity, we
consider the correction function in dimensionless form
$F_{\rm e}=g^4(1/g^2-g^2)/y^2$. Since the solution of Equation~(\ref{dgdtb})
relates $g$ and $y$, $F_{\rm e}$ is, in fact, a function of one variable. In
Figure~\ref{cac}, we show it as a function of $g$ (left panel) and as
a function of the variable $\k=t(1-V_{\rm c})/t_0=2D(y-1)\Gamma_{\rm
  j}^2(1-V_{\rm c})$ (right panel), corresponding roughly to the
observation time (here $V_{\rm c}=\sqrt{1-\Gamma_{\rm c}^{-2}}$ is the
blob velocity). The values of the parameter $D=0.1$, $1$, $10$ and
$100$ are used in both panels of Figure~\ref{cac}. It can be seen that in
the case of $D\gtrsim 1$, the solution depends weakly on the parameter
$D$, and the non-thermal activity of the blob is expected to have a
narrow peak of duration:
\begin{equation}
t_0\approx\frac{z_0}{c}\frac{1}{D}\frac{1}{2\Gamma_{\rm j}^2}.
\label{dtpeak}
\end{equation}
The maximum of the correction function occurs at $g_{\max}\approx 0.8$.

In the case of $D\ll 1$, the situation is quite different. Namely, the
expected nonthermal activity has no pronounced peak, thus such a blob
cannot produce a high-amplitude flare. In this regard, we can
formulate the condition $D\gtrsim1$ as a requirement for a flaring
episode. This condition can be reformulated as a upper limit on the
cloud mass:
\begin{equation}
M_{\rm c,rc}\lesssim \frac{L_{\rm j} r_{\rm c}^2}{4\theta^2 c^3 z \Gamma_{\rm j}^3}\,.
\label{mcrc}
\end{equation}
The above relation depends not only on the properties of the cloud
(its size and mass) but also on the jet power and Lorentz factor.
Thus, for quantitative calculations, one needs
detailed information about the dynamics and properties of the blazar
jet.

Although the process of jet formation is not fully understood,
 recent hydrodynamical studies of different scientific groups show that the
Blandford-Znajeck \citep{ruffini75,lovelace76,BZ77} process may be at work in AGN, and suggest a concept of
magnetically accelerated jets. Thus, the jet base is expected to be
strongly magnetized and likely magnetically dominated at $z\le 1
\mbox{ pc}$ \citep{kbvk07,bk08,bes10}. This immediately gives the
magnetic field strength of the jet in laboratory frame
\begin{equation}
B_{\rm j}\approx \left(\frac{4 L_{\rm j}}{c z^2 \theta^{2}}\right)^{1/2}
\approx 120\,L_{\rm j,46}^{1/2}z_{17}^{-1}\theta_{-1}^{-1}\,\mbox{G}\,.
\label{bj}
\end{equation}
During the jet propagation, the magnetic field energy can be transformed
to the bulk kinetic energy. At the linear stage a simple relation defines the bulk
Lorentz factor \citep{bn06}
\begin{equation}
\Gamma_{\rm j}\approx \frac{\omega}{4 r_g}\,.
\label{gam_m}
\end{equation}
Finally, the opening angle of the jet is
expected to be $\theta\approx 1/\Gamma_{\rm j}$ \citep{kvkb09}.
Combining Equations~(\ref{bj}) and (\ref{gam_m}), one can
estimate the magnetic field in the jet comoving frame,
\begin{equation}
B_{\rm c}\approx \frac{2}{z}\left(\frac{L_{\rm j}}{c}\right)^{1/2}\approx 12
z_{17}^{-1} L_{\rm j,46}^{1/2}\,\mbox{G}\,.
\label{bjc}
\end{equation}
In Figure~\ref{magjet}, the  typical magnetic field and bulk Lorentz factors
of the jet are shown for  three different distances from the BH. These values are in good agreement
with observed values of magnetic field on parsec scales in AGNs \citep{lob98,swvt08,sg09,skl10}.

Using Equation~(\ref{gam_m}), one can present Equation~(\ref{mcrc}) in the form
\begin{equation}
M_{\rm c,rc}\lesssim 0.5\times10^{26} L_{\rm j,46}r_{\rm c,15}^2 \Gamma_{\rm j,1.5}^{-3}M_{\rm BH,8}^{-1}\,\rm g.
\label{mcrcg}
\end{equation}
Note that at this late stage, the cloud can be already significantly
expanded with a radius of $r_{\rm c,15} \gg 1$.
The extreme value of $M_{\rm c,rc}$ can be achieved at $r_{\rm c}\approx \omega$:
\begin{equation}
M_{\rm c,rc}\lesssim 2\times10^{26} L_{\rm j,46} M_{\rm BH,8}^{} \Gamma_{\rm j,1.5}^{-1}\,\rm g.
\label{mcrce}
\end{equation}
This upper limit is more robust than the constraints given by
Equations~(\ref{dms}),(\ref{mcsc}), and (\ref{mcjc}), we note however, 
that it concerns the mass of the blob, but not the mass of the ablated stellar atmosphere.

\subsection{Energy Budget of the Cloud}

Below we consider the general requirements to the scenario in the
context of blob's radiation efficiency. These constraints
have a quite basic character and are not related to a specific radiation
mechanism. Obviously, the gamma-ray production mechanisms impose
additional requirements, concerning e.g. the density of relevant
targets in the form of gas, radiation or magnetic field. We discuss the
impact of specific radiation mechanisms in
Section~\ref{sec:proc}. Here we try to find a generic link between
properties of a blob, which is responsible for the nonthermal emission,
to the parameters of the AGN, i.e. the mass of the central engine and the jet power.

Emission detected from blazars are significantly enhanced
due to Doppler boosting.  The apparent  ($L_\gamma$) and intrinsic
($L_{\rm sc}$) luminosities are connected through the well known
relation: $L_{\gamma}=L_{\rm sc}\delta_{\rm c}^4$ (here $\delta_{\rm
  c}$ is Doppler factor of the production region). On the other hand,
the intrinsic luminosity of the blob can be expressed as a fraction
$\xi$ of the power transferred by the jet to the blob:
\begin{equation}
L_{\rm sc}=\xi \left(\frac{1}{\Gamma_{\rm c}^2}-\frac{\Gamma_{\rm c}^2}{\Gamma^4_{\rm j}}\right)\frac
{L_{\rm j} r_{\rm c}^2}{4\omega^2 } \, .
\label{eq:lsc}
\end{equation}
The parameter $\xi\ll1$ accounts for the overall efficiency of of transformation of the
absorbed jet energy to nonthermal emission through acceleration and radiation of
relativistic particles (note that the righthand-side of the
Equation~(\ref{eq:lsc}) refers to the quantities in the observer frame,
whilst the lefthand-side corresponds to the blob's reference frame). 
For small aspect angles, $\delta \approx 2\Gamma_{\rm c}$, we thus 
obtain the following simple relation:
\begin{equation}
L_{\gamma}=4\xi F_{\rm e} L_{\rm j} \Gamma_{\rm j}^2 \frac
{ r_{\rm c}^2}{\omega^2} \ ,
\label{eq:lgamma}
\end{equation}
which has a few interesting implications. In
particular, one can estimate the size of the blob:
\begin{equation}
 r_{\rm c}=\frac{\omega}{\Gamma_{\rm j}} \left({ L_{\gamma}\over 4 \xi F_{\rm e} L_{\rm j}}\right)^{1/2}\,.
\label{eq:blob_size}
\end{equation}
Given the standard bulk Lorentz factors, $\Gamma_{\rm j}\approx \omega/(4r_{\rm g})$, 
and the maximum of correction function, $\max{(F_{\rm e})}\approx0.4$, together with 
conventional normalizations, one obtains
\begin{equation}
r_{\rm c}\approx  5\times 10^{14} M_{\rm BH,8} L_{\gamma,47}^{1/2} L_{\rm j,46}^{-1/2}
\xi_{-1}^{-1/2}\, \mbox{ cm}.
\label{rce}
\end{equation}
Here we use quite a high normalization for the
efficiency, of $\xi_{-1}=\xi/0.1$

Another important estimate can be obtained for the
maximum apparent luminosity of the blob.
It is achieved when the blob eclipses the whole jet,
i.e. $r_{\rm c}\approx \omega$. In this case, one obtains:
\begin{equation}
L_{\gamma\max}=2\times10^{48}\xi_{-1} L_{\rm j,46}\Gamma_{\rm j,1.5}^2\, \rm erg\,s^{-1}.
\label{eq:max_lum}
\end{equation}
where $\Gamma_{\rm j,1.5}=\Gamma_{\rm j}/10^{1.5}$.
Note that the apparent non-thermal
luminosity of the blob is proportional to $\Gamma_{\rm j}^2$. Remarkably,
even for relatively modest values of the jet Lorentz factor, $\Gamma_{\rm j} \sim 10$,
and the conversion efficiency, $\xi \sim 0.01$, $L_\gamma$ is comparable to the
jet power.

\subsection{Time variability}
The fast variability of TeV gamma-ray emission of blazars, which can
be as short as a few minutes as reported for \pks{}
\citep{ah07pks} and Mkr~501 \citep{mkr501_magic}, is a key
observational fact which should be addressed by any model of TeV
blazars.  In addition to a general (standard) statement about the
strongly {\it Doppler boosted} gamma-ray emission produced in very
{\it compact regions} close to the central black hole, any dedicated
model should provide intrinsic reasons for variability (characterizing
the scenario as a whole) and offer radiation mechanisms with adequate
cooling times. If the cooling time-scale appears to be too long, an
alternative source of the variability is required. For example, a
change of the production site velocity may lead to a strong change in
the apparent luminosity. Indeed, since the production region is to be
Doppler boosted, a variation of the Doppler factor may be a plausible
reason for strong variation of the observed flux.  

In this section, we
discuss the variability scales related to this effect, and its
implication to the JRGI scenario. We note that the relevant
time-scales do not depend on the cooling time of the emitting
particles. In particular, the obtained size of the blob
in Equation~(\ref{rce}) and the jet Lorentz factor provide following 
lower limit on the variability time-scale:
\begin{equation}
\tau>\frac{r_{\rm c}}{\Gamma_{\rm j} c}\approx4\times10^2z_{17}^{-1/2}L_{\gamma,47}^{1/2}L_{\rm j,46}^{-1/2}\xi_{-1}^{-1/2}M_{\rm BH,8}^{3/2}\,\rm s\,,
\label{size_variability}
\end{equation}
which appears to be close to the observed one, and can be
significantly shorter in the case of powerful jets.

\subsubsection{Duration of the blob-jet interaction}
The principal variability scale in the JRGI scenario is related to the duration of
the effective interaction of the cloud with the jet and is determined by the 
function $F_{\rm e}$ (see Section~\ref{acac}).
Since the model requires very effective acceleration of particles, with
a $\geq 10 \%$ efficiency of the energy transformation to nonthermal particles,
the shape of the function, $F_{\rm e}$, (see Figure~\ref{cac}) can be treated as the
time profile of particle acceleration with a characteristic timescale:
\begin{equation}
\Delta t\approx\frac{2z_0^2\theta^2}{r_{\rm
c}^2}\frac{\Gamma_{\rm j}c^2 M_{\rm c}}{L_{\rm j}} \ .
\label{eq:time_scale}
\end{equation}
Note that in the extreme case, when the blob eclipses the entire jet, i.e.
$z_0^2\theta^2/r_{\rm c}^2 \sim 1$, the characteristic time-scale, $\Delta t$, depends only on the
jet Lorentz factor $\Gamma_{\rm j}$ and power $L_{\rm j}$, as well as on the mass
of the cloud $M_{\rm c}$:
\begin{equation}
\Delta t\approx 60 \Gamma_{\rm j,1.5}L_{\rm j,46}^{-1}M_{\rm c,25}\, \rm s \ ,
\label{eq:time_scale_n}
\end{equation}
The total apparent energy of electromagnetic radiation which can be emitted by the cloud
can be estimated from Equations~(\ref{eq:max_lum}) and (\ref{eq:time_scale_n}):
\begin{equation}
E_{\rm tot}\approx 10^{50}\xi_{-1} M_{\rm c,25} \Gamma_{\rm j,1.5}^3\,\rm erg.
\label{eq:e_tot}
\end{equation}

For the values expected in this scenario (which were also used for
normalization of $\Gamma_{\rm j}$, $L_{\rm j}$ and $M_{\rm c}$ in
Equation~(\ref{eq:time_scale_n})), the cloud-jet interaction can be quite brief;
- shorter than the detected variability of the ultrafast flares of
\pks{} and \mkn501.  Moreover, for small mass clouds, it can
be as short as 1~sec. Obviously, this time-scale corresponds to the
flare rising interval, while fast emission decay requires short
radiative or adiabatic cooling of the emitting particles or rapid changes
in the blob's Doppler factor.

\subsubsection{Helical structure of relativistic jet}

Generally,  in powerful jets, the change of the Doppler factor 
is unavoidable. Indeed, since the matter in the jet moves along the dominant
magnetic field lines, which are expected to be helical, the velocity
of the blob should have both poloidal and azimuthal components. The
azimuthal velocity can be as high as $v_{\phi}\approx c r_{lc}/\omega$
\citep{bn06,kbvk07,bes10}, where $r_{lc}\sim4r_{g}$ is light cylinder radius, 
therefore, during the motion, the angle towards the
direction to observer can be changed. We discuss the corresponding
variability pattern in Appendix~\ref{aboost}, where it is shown that this
effect (which reminiscent to some extent, the operation of a `revolver'), can
lead to a change in the flux by a factor of 2 if the blob has 
$v_{\phi}\sim 0.5 c$ and turns around the jet axis by an angle of $\sim\pi/4$. 
In the case of magnetically driven jet using Equation~(\ref{gam_m}) we can get 
$v_{\phi}\approx c /\Gamma_{\rm j}$ or if $\Gamma_{\rm j}>3$ this mechanism cannot
to explain variability of blazars.


\subsubsection{Collision of blobs}

The interaction of the jet with a massive cloud leads unavoidably to
the formation of a large number of small blobs, which may gain 
an additional chaotic velocity component.  Let us assume that the chaotic
velocity is comparable to the sound speed in  relativistic gas, i.e. $v_{\rm
  s} \sim c/\sqrt{3}$, which is larger than $c/2$, that is enough to explain 
strong radiation variability (see for details Appendix~\ref{aboost}). 
Due to interactions with each other, the blobs
can change their speeds on timescales $t_{\rm s}\sim 2 r_{\rm
  c}/v_{\rm s}$ leading to the variability on a timescale
\begin{equation}
\tau \approx t_{\rm s}/\delta \gtrsim 5 \times 10^{2} M_{\rm BH,8} L_{\gamma,47}^{1/2} L_{\rm j,46}^{-1/2} \xi_
{-1}^{-1/2}\delta_2^{-1} \, \rm s,
\label{lfv}
\end{equation}
where Doppler factor is determined as $\delta_2=\delta/10^2$.

Equation~\eqref{eq:time_scale_n} shows that the interaction time-scale
in the JRGI scenario could be very short allowing, in the case of
comparably short particle cooling time, flaring episodes of duration
$\sim 100$~s. If the particle cooling mechanism cannot provide the
required energy loss rate, the nonthermal flux variability can be
caused by a change in the production region Lorentz factor, see
e.g. Equation~\eqref{lfv}. However, we note that
in this case the variability on timescales as short as 100~s would require a
rather specific combination of several principal parameters.
Thus, a radiative mechanism with short
cooling time remains still a quite feasible requirement for  models
intended to explain the fast variability observed in blazars.  

In the
following section we discuss the efficiency and features of major
radiation mechanisms related to both protons and electrons.  Although
the jet composition is still debated, the conventional approach
attributes the non-thermal activity of AGN to a lepton IC mechanism. In the
suggested scenario even if the primary content of the jet is leptonic,
the ablated cloud itself may provide protons for the acceleration
process.  In this paper we do not discuss the specific mechanisms of
particle acceleration, but simply assume that both electrons and
protons are effectively accelerated during the interaction of the blob
with the jet.

\section{Radiation mechanisms}
\label{sec:proc}

In this section we discuss the applicability of different radiation
mechanisms responsible for the gamma-ray emission of blazars in the
context of the shortest variability timescale of order of 100~s
observed during strongest flares of \pks.
The spectral energy distribution (SED) of the source has a typical
shape for such objects with two pronounced humps. The low energy
component peaks in the optical-UV-soft X-ray band at $\nu
\sim10^{16} \mbox{ Hz}$, whilst the high energy bump has a maximum in the
VHE band. The location of the gamma-ray maximum is measured only in
the low state of the source, during simultaneous Fermi-HESS
observations in 2009, which revealed a broad maximum between 10~GeV and
100~GeV \citep{a_pks_low_09}.  Although the exact position of the
gamma-ray maximum is not yet measured in the high state of the source,
the observed spectral flattening  during the July
2006 flares \citep{ah_pks_09} indicates a tendency of the extension of the
region of the flux maximum, but most likely not far beyond
100~GeV. Note that for the distance to the source of $D=540\rm Mpc$
($z=0.116$), the attenuation of gamma-rays of energy $\ll 1$~TeV in
the extragalactic background light (EBL) is negligible. Therefore
throughout this paper we will assume that the gamma-ray peak in the
SED of this source is located around $100$~GeV.  The average apparent
gamma-ray luminosity of the 2006 July flares was at the level of $\sim
10^{47} \ \rm erg\,s^{-1}$. During the giant July 28 flare, the source was
not monitored in the X-ray energy band, but the simultaneous
observations of the next night, also characterized by strong flares,
conducted with the H.E.S.S., Chandra and Bronberg optical telescope,
revealed that the luminosity of the object in optical, UV and X-ray
energy bands was  an order of magnitude lower compared to the
gamma-ray luminosity.  These rather general properties of the SED,
which include peak location and flux ratio, allow us to derive some
important constraints on the production mechanisms.

\subsection{Gamma rays associated with electrons}

\subsubsection{SSC model}
\label{sec:ssc}
In the SSC models the high energy gamma-rays are produced by
relativistic electrons through IC scattering of synchrotron radiation
of the same electron population. Generally, this model satisfactorily
explains the basic features of gamma-ray blazars. However, the ultrafast
flares of \pks{} pose severe constraints on the parameters
characterizing the gamma-ray production region. Generally, the IC
scattering proceeds in the Thomson regime, when $h \nu E_{ \gamma}\ll
m^2c^4\delta^2$, where $\delta\gg1$.  Thus, in the co-moving reference
frame,
\begin{equation}
E_{\gamma,11}=4\times10^{-10}\nu_{16} \gamma^2,
\label{eq0}
\end{equation}
where $\gamma$ is the electron Lorentz factor,
$E_{\gamma,11}=E_{\gamma}/100\rm GeV$ and
$\nu_{16}=\nu/10^{16}\rm Hz$ are the peak energies of IC and synchrotron
components in the observer reference frame.  Then, the electron Lorentz
factor can be estimated as
\begin{equation}
\gamma=5\times10^4 \left(\frac{E_{\gamma,11}}{\nu_{16}}\right)^{1/2}.
\end{equation}

The strength of the magnetic field in the co-moving frame can be defined  from the
 location of the synchrotron peak,
\begin{equation}
\nu = 6\times10^6 B_{0}\gamma^2\delta ~ \rm Hz,
\label{eq1}
\end{equation}
where $B_{0}=B/1 \rm G$. Here, for order-of-magnitude estimates, we
adopt that the maximum of the $\nu F_\nu$ distribution of synchrotron
photons occurs at energy $1.33 \omega_{\rm c}$\footnote{This energy is
in fact used in Equation~(\ref{eq0}) for IC scattering in the Thomson
regime}, 
where $$\omega_{\rm c}=\frac32\frac{eBE^2}{m^3c^5}$$ is the synchrotron characteristic
frequency, while the maximum of the $F_\nu$ distribution  is 
located at a lower energy of $0.29 \omega_{\rm c}$.  
Thus, one obtains:
\begin{equation}
B_{0}=0.7\,{\nu_{16}^2}{E_{\gamma,11}^{-1}\delta^{-1}}.
\label{bssc}
\end{equation}

The ratio of the IC and synchrotron peak luminosities, $f$, is another
important parameter characterizing two-hump SEDs:
\begin{equation}
 f=\frac{L_{\rm IC}}{L_{\rm syn}}=\frac{w_{\rm ph}}{ w_{\rm B}}.
\label{eq6}
\end{equation}

In the SSC model, taking into account the constraint on the size of the production
region imposed by the observed variability of the time-scale in 
Equations~(\ref{eq_size_prop}--\ref{db}), a lower limit for
the co-moving energy density of the target photons can be obtained:
\begin{equation}
w_{\rm ph} \ge 10^{10} \frac{L_{\rm X,46}}{ \tau_{2}^2\delta^6} ~\mbox{ erg cm}^{-3},
\label{eq7}
\end{equation}
where {$L_{X,46} = L_X/10^{46}$ erg s$^{-1}$} is the apparent synchrotron luminosity. For the given luminosity ratio of the observed high and low energy
components, one finds
\begin{equation}
\delta \approx 900 \left( \frac{L_{\rm X,46}E_{\gamma,11}^2}{f
\tau_{2}^2\nu_{16}^4}\right)^{1/4} \, .
\label{delssc}
\end{equation}
Thus, for SEDs, typical for ultrafast flares of \pks, the
Doppler factor of the relativistically moving gamma-ray source should
exceed $\delta\sim500$.  Note that this condition is stronger than the
constraint on the Doppler boosting imposed by the condition of the
gamma-ray transparency of the source \citep{bfr08}.

The order-of-magnitude estimates of Equations~(\ref{bssc}) and (\ref{delssc}),
obtained for SSC scenario, suggest that short flares from \pks{}
should be produced at large distances from the BH (given the weak magnetic
field and large Doppler boosting factors). This requirement is, in
fact, very constraining for JRGI scenario, since the jet ram pressure
in this region appears to be extremely small, viz.  
\begin{equation}
P_{\rm ram, SSC}\approx\frac{B_0^2\Gamma_{\rm j}^2}{8\pi}\approx 5\times
10^{-3}\nu_{16}^4E_{\gamma,11}^{-2}\,\rm dyn, cm^{-2}\,,
\label{pram_ssc}
\end{equation}
which is not enough (by far) to ablate the required amount of stellar material.

The above estimates show that SSC models meet severe limitations in
the framework of JRGI scenario due to the required weak magnetic
field. 
We note that the above severe constraints are basically due to the very small 
magnetic field. In principle, one can assume that the magnetic field in the blob 
is much weaker then in the jet, which could improve the effect of 
the SSC mechanism. However, in any case, Equation~(\ref{delssc})  
can be satisfied only at very large distances, where the
jet ram pressure is very small and to oblate and accelerate the stellar atmosphere is difficult.

\subsubsection{Model of external photon field}

The main difference between EIC and SSC models is that in the former
one the gamma radiation is dominated by the scattering of electrons on
low-energy photons of external origin.  Obviously, in such a case the
energy density of the external photon field should exceed significantly the energy
density of the synchrotron photons, i.e. the photon energy density in
the jet vicinity should fulfill the following requirement
\begin{equation}
 w_{\rm ext} \ge 10^{10} \frac{L_{X,46}}{ \tau_{2}^2\delta^6\Gamma_{\rm j}^2} ~\mbox{ erg cm}^{-3},
\label{photon_ext}
\end{equation}
which follows immediately from Equation~(\ref{eq7}). In fact, even a very
weak external photon field can fulfill this requirement, given a very
strong dependence on the jet Lorentz factor.  Another limitation on
the external photon field can be derived from the ratio of the IC and
synchrotron peaks (see Equation~(\ref{eq6})) which together with
Equations~(\ref{gam_m}) and (\ref{bjc}) give the following luminosity limit:
\begin{equation}
L_{\rm ext}\ge3\times10^{42}fL_{\rm j,46}z_{17}^{-1}M_{\rm BH,8}\,\rm erg\,s^{-1}\,,
\label{external_lum}
\end{equation}
which is likely available  in the vicinity of powerful blazars.   Since this photon field
may remain undetectable, the energy of the target photon is,
to a large extent, a free parameter. Thus, Equation~(\ref{eq0}) is
not valid in the EIC case, and the model parameters are less constrained
than in the SSC scenario. In particular, this allows us to relax the
requirement of weak magnetic field, which is crucial for
SSC models in the framework of JRGI scenario. To make 
quantitative estimates, one needs to assume some basic properties of
the external field, namely the typical photon energy $\epsilon_{\rm
  eV}=\epsilon/1~\rm eV$ and its luminosity $L_{\rm ext}$. Then
Equations~(\ref{eq0}) and (\ref{bssc}) obtain the following form:
\begin{equation}
E_{\gamma,11}= 10^{-11}\epsilon_{\rm eV}\gamma^2\delta^2\,,
\label{eq0_eic}
\end{equation}
and
\begin{equation}
B_{0}=2\times10^{-2}\nu_{16}\epsilon_{\rm eV}\delta E_{\gamma,11}^{-1}\,.
\label{bssc_eic}
\end{equation}
Together with Equations~(\ref{gam_m}) and (\ref{bjc}), one can solve
these equations and represent $\gamma$ and $\epsilon_{\rm eV}$ as
\begin{equation}
\epsilon_{\rm eV}\approx10L_{\rm j,46}^{1/2}E_{\gamma,11}M_{\rm BH,8}^{1/2}\nu_{16}^{-1}z_{17}^{-3/2}\,,
\label{epsilon_eic}
\end{equation}
and
\begin{equation}
\gamma\approx1.5\times10^3\nu_{16}^{1/2}z_{17}^{1/4}M_{\rm BH,8}^{1/4}L_{\gamma,46}^{-1/4}\,.
\label{gamma_eic}
\end{equation}
An important characteristic of the model is the emitting particle
cooling time in the jet reference frame:
\begin{equation}
t_{\rm cool}'= 3\times10^7 w_0^{-1} \gamma^{-1} \; \rm s\,,
\label{eq01o}
\end{equation}
where $w_0=w_{\rm B}+w_{\rm ph}=(f+1)w_{\rm B}$ is the comoving frame energy density
of the target fields in units of $\rm erg\, cm^{-3}$ . Equations~(\ref{bjc}) and (\ref{gamma_eic}) 
result in the following cooling time:
\begin{equation}
t_{\rm cool}'= {3\times10^3}{\left(1+f\right)^{-1}}z_{17}^{7/4}L_{\rm j,46}^{-3/4}M_{\rm BH,8}^{-1/4}\nu_{16}^{-1/2}\,\rm s\,.
\label{eq01}
\end{equation}
This fast cooling allows us to relate the observed variability on timescale of several
hundred seconds, with the relativistic particle radiation cooling. 
We note as well that given
the relatively low energies of the electrons, $E_{\rm e}\sim1-10$~GeV, and
the strong magnetic field ($B\sim$ few G), the acceleration of these electrons can be
easily realized.

Finally, we have to note that the obtained values should fulfill the
condition of IC scattering in the Thomson regime: $\gamma\epsilon_{\rm
  ext}\Gamma_{\rm j}\ll mc^2$ (since some used relations,
e.g. Equation~(\ref{eq0_eic}), are valid in the Thomson regime only). This
yields the following requirement for the interaction point:
\begin{equation}
z_{17}\gg L_{\rm j,46}^{1/3}M_{\rm BH,8}^{1/3} E_{\gamma,11}^{4/3}\nu_{16}^{-2/3}\,.
\label{z_eic}
\end{equation}
In the case that this condition is not true, the interaction of electron
with target photons occurs in the Klein-Nishina regime and a few
additional effects, such as Klein-Nishina electron losses and gamma-gamma
attenuation, have to be taken into account. Actually, since in this
case, the Klein-Nishina losses have to be the dominant cooling mechanism,
the gamma-gamma absorption is unavoidably large. Indeed, the Klein-Nishina cooling time is
\begin{equation}
t_{\rm KN}'=\frac{E}{\dot{E}_{\rm KN}}\approx\frac1{cn_{\rm ext}'\sigma_{\rm KN}}\approx 10^2\tau_2\Gamma_{\rm j}\,\mbox{ s},
\label{kn_cooling}
\end{equation}
where $n_{\rm ext}'$ and $\sigma_{\rm KN}$ are the target photon
density in the jet frame and Klein-Nishina cross-section,
respectively. On the other hand, the gamma-gamma optical depth can be estimated as 
\citep[see][]{der09}
\begin{equation}
\tau_{\gamma\gamma}=zn_{\rm ext} \sigma_{\gamma\gamma}\approx
40M_{\rm BH,8}\tau_2^{-1}\,,
\label{kn_tau}
\end{equation}
where $n_{\rm ext}$ and $\sigma_{\gamma\gamma}$ are the target photon
density in the laboratory frame and pair-production cross-section. To
derive Equation~(\ref{kn_tau}), we have used the approximate relation
$\sigma_{\gamma\gamma}\approx2\sigma_{\rm KN}$ and Equations~(\ref{gam_m}) and
(\ref{eq01}). Thus, it is rather unlikely that short flares can be
produced in the Klein-Nishina regime on the external photon field.

\subsection{Gamma rays associated with protons}

\subsubsection{pp and p$\gamma$ interaction}

The production of gamma-rays from interactions of relativistic protons with
the surrounding gas is one of the major processes in high energy
astrophysics.  This process is effective in relatively dense
environments, namely when the pp cooling time, $t_{\rm
  pp}=10^{15}/n$~s does not exceed other characteristic times (here
$n$ is the target proton density in units of $\rm cm^{-3}$). In the case of the blob
in the jet, the most relevant times are the interaction time of the
blob with the jet given by Equation~(\ref{eq:time_scale_n}) (which can be
treated as the acceleration time of protons), and the escape time of
protons from the blob. The number density of protons in the blob is
constrained by Equations(\ref{mcrce}) and (\ref{rce}). For the parameter
values expected in the scenario, the number density of protons does
not exceed $10^4 \ \rm cm^{-3}$.  The corresponding {\it pp} cooling
time of $\geq 10^{11} \ \rm s$ is too long, and does not leave any room
for the explanation of the variability of TeV radiation on any observed
time-scale. Even the assumption that the variability is caused by other
reasons, e.g. due to the adiabatic cooling or change of the Doppler
factor, cannot help much since in this case, an extremely low
efficiency represents an unavoidable argument against this process.

The efficiency of gamma-ray production can be much higher through other channels
related to interactions with the radiation and magnetic fields. In both cases
the gamma-ray production rate increases dramatically with the energy of protons,
and achieves reasonably high efficiency if the protons are accelerated to
energies of $10^{19} \ \rm eV$ or beyond. For the compact blobs with linear dimensions
severely constrained by the hour-long  or shorter variability timescales,
the energy of protons can achieve such high energies only when the
particle acceleration proceeds (1) at a rate close
to the theoretical limit, $t_{\rm acc} \sim r_{\rm L}/c$ and
(2) in the presence of a magnetic field
as large as 100~G \citep{ah00}. Whilst for these conditions both the acceleration and
synchrotron cooling times in the frame of the blob can be as short as 1h,
the fast cooling of protons via photomeson interactions require very dense photon
fields at mm and far-infrared wavelengths. On the other hand, the density of the
radiation field (of internal or external origin) is constrained by the condition of
transparency of the production region for  the VHE gamma-rays, implying that
the optical depth regarding the photon-photon pair production cannot
significantly exceed unity. This condition, coupled with the condition
of a large magnetic field amplitude of $B \sim 100$~G, makes the cooling time of protons
via photomeson processes significantly longer compared to the proton
synchrotron cooling time \citep{ah00}. Although, formally one can ``construct"
a model with extreme parameters, where the photomeson processes could compete
with the proton synchrotron cooling, below we will focus our treatment on the
production of gamma-rays via synchrotron radiation.

\subsubsection{Proton synchrotron radiation}

Protons of extremely high energy and large magnetic fields strengths are the two
conditions which make proton synchrotron an effective radiation
mechanism. If these conditions are satisfied, the spectrum of
synchrotron radiation can extend to the gamma-ray domain with a
characteristic energy \citep{ah00}:
\begin{equation}
E_{\gamma,11}
\approx 1\, B_{2}E^2_{19}\,,
\label{eq21}
\end{equation}	
 where $E_{19}=E/10^{19} \rm eV$ is the proton energy, and $B_{2}=B/100 \; \rm G$
 is the strength of the magnetic field.
The position of the peak depends strongly on the maximum energy of protons,
which is determined by the balance between the particle
acceleration and cooling rates. It is convenient to present the
acceleration time of the protons, independent of the specific
mechanism of acceleration, in the form:
\begin{equation}
t_{\rm acc}=\frac{\eta(E) r_{\rm L}}{c}\approx10^4 E_{19}B_{2}^{-1}\eta(E) ~\rm s \ ,
\label{acctime}
 \end{equation}
where $r_{\rm L}=E/eB=3\times10^{14}B_2E_{19}\,\rm cm$ is the so-called gyro-factor. 
We note that this value is remarkably  close to the required size of the blob in the 
JRGI scenario (see Equation~(\ref{rce})). The dimensionless parameter, $\eta(E) \ge 1$, in 
Equation~(\ref{acctime}) characterizes the acceleration efficiency; the most 
efficient acceleration occurs for $\eta=1$.
It is believed that the acceleration by relativistic shocks \citep[see e.g.][]{agk01}
or at the annihilation of the magnetic field lines \citep{hts92} occurs in
the regime when $\eta \sim 1$. In a more general context, the relativistic outflows found
in AGN and GRBs, which presently are considered as most feasible sites of the production 
of the highest energy cosmic rays, can accelerate protons to $10^{20} \ \rm eV$ only if
$\eta \sim 1$ \citep{abd02}.
In the comoving frame, the characteristic synchrotron cooling time
of the protons is:
\begin{equation}
t_{\rm sy}= \frac{4\pi m_{\rm p}^4c^3}{ \sigma_{\rm T} m^2_{\rm e}EB^2}=3 \times 10^4
B^{-2}_{2}E^{-1}_{19} ~ \rm s.
\label{cooltime}
\end{equation}
From comparison of the acceleration and cooling rates one finds a
maximum energy of protons of;
\begin{equation}
E_{\rm p, max}=1.7\times10^{19}B_{2}^{-{1/ 2}}\eta^{-{1/ 2}}~\rm \mbox{ eV} \ ,
\label{eppeak}
\end{equation}
and a corresponding maximum energy of synchrotron photons of;
\begin{equation}
E_{\gamma, \max}
= 400 \eta ^{-1}
~\rm \mbox{ GeV} \ .
\end{equation}
If synchrotron radiation is emitted by a blob with a Doppler-factor, $\delta$,
the peak of the synchrotron radiation is shifted to
\begin{equation}
E_{\gamma, \max} \approx 400 \eta ^{-1}\delta ~\rm GeV \ .
\label{egpeak}
\end{equation}
One can see that for a Doppler factor of $\delta \le 100$, 
synchrotron radiation can extend to TeV energies provided that the
proton acceleration rate is close to $\eta \le 100$. 

Another important constraint can be obtained through so-called 
{\it Hillas criterion}, i.e. a requirement for the size of the
acceleration site to be larger than the gyroradius of the highest
energy particles.  Substituting
Equations~(\ref{bjc}) and (\ref{eppeak}) into the Larmor
radius, $r_{\rm L}\approx {E}/{eB_{\rm c}} $, one obtains
\begin{equation}
z_{17}^{3/2}L_{\gamma,47}^{-1/2} L_{\rm j,46}^{-1/4} \eta_1^{-1/2} \xi_{-1}^{1/2}M_{\rm BH,8}^{-1} <0.1 \; .
\label{psca}
\end{equation}
Here the size of the acceleration site was assumed to be defined by
Equation~(\ref{rce}).  A significantly more severe constraint for the
proton synchrotron models comes from the cooling time
requirement. Namely, the expected variability time scale can be 
estimated from Equations~(\ref{gam_m}), (\ref{bjc}), (\ref{cooltime}) and (\ref{eppeak}):
\begin{equation}
  \tau_{\rm psyn}\approx\frac{t_{\rm sy}}{\d}\approx 2\times10^4 \eta_1^{1/2}
M_{\rm BH,8}^{1/2} z_{17}L_{\rm j,46}^{-3/4}\, \rm s\,.
\label{syn_pr_cooling}
\end{equation}
Thus, in the case of powerful jets, $L_{\rm j}>10^{47}\,\rm erg\,s^{-1}$, 
the proton synchrotron mechanism can guarantee variability on time-scales of several hundred seconds.


Although a detailed study of broad-band SED in the framework of
proton synchrotron scenario remains outside of the scope of this
paper, below we outline some expected features. In
particular, one may expect that the acceleration of protons is
accompanied by the acceleration of electrons, whose particle population
may create a detectable nonthermal emission component.  Due to more
effective synchrotron losses, the maximum energy of electrons is
rather small:
\begin{equation}
E_{\rm e}= 5 B_{2}^{-1/ 2}\eta_{\rm e}^{-1/ 2} \rm ~ TeV \ .
\end{equation}
The corresponding synchrotron peak appears at an energy, which is by a factor of
$m_{\rm p}/m_{\rm e}$ smaller than the peak energy of proton  synchrotron (see Equation~(\ref{egpeak})):
\begin{equation}
 E_{\rm syn}\approx150 \eta_{\rm e}^{-1} \rm ~MeV.
\end{equation}
Thus, in this scenario, the synchrotron peak of electrons,
additionally boosted by the value of the Doppler factor,
$\delta\sim50$, is expected at GeV energies, but not in the
traditional X-ray radiation band. In this regard the so-called
synchrotron peak, which in TeV blazars is located in the UV to X-ray
band,  needs a new (non-standard)
interpretation in the framework of our model. Finally, it is important to note that the flux ratio
of proton to electron synchrotron peaks depends on the ratio of the
injection rates of those particles, which is an highly uncertain
parameter. 

The peak of soft radiation in the region of $10^{16} ~\rm Hz$ can be produced
by secondary electrons produced in interactions of VHE gamma-rays with
soft photons. A modest absorption of VHE gamma-rays
cannot be excluded. Moreover, in the case of blazars with extremely hard
TeV source spectra (i.e. after correction for the intergalactic absorption),
the energy-dependent absorption can be a natural explanation of
the unusually hard gamma-ray spectra \citep{akc08,zck10}.

The energy of the absorbed gamma-ray photon is shared between an
$e^{\pm}$ pair, so each electron on average gets the half of the
original energy, i.e., $E_{\rm e} \sim E_{\gamma}/2$. Generally, synchrotron
radiation of secondary electrons produced inside the jet will be
Doppler boosted, therefore this amplified component of synchrotron
radiation can be detected by the observer.  The peak energy of the
secondary synchrotron radiation is quite sensitive to the Doppler
factor of the blob and the energy spectrum of the parent gamma-rays. For
the standard values used in this paper, the peak of the
synchrotron radiation can vary in a quite broad interval between UV
and hard X-ray \citep{zck10}.

\begin{figure}
\plotone{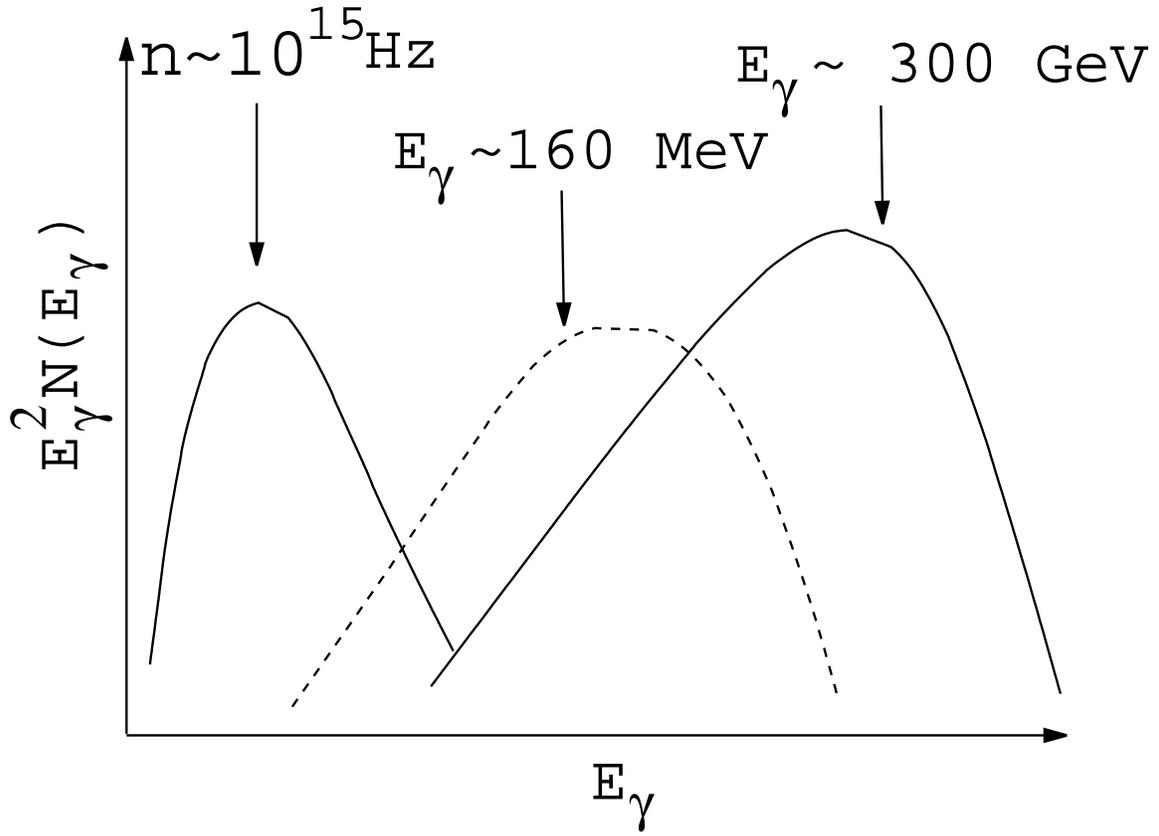}
\caption{Expected SED of blazars in the framework of the
 proton-synchrotron scenario in the blob's reference frame. The maximum at
 $10^{15} \;\rm Hz$ is due to synchrotron radiation from secondary (pair-produced) electrons; the maximum at $\sim
 100 \rm MeV$ corresponds to the
 synchrotron radiation of primary electrons accelerated in the
 blob; the maximum at $ \sim 300 GeV $ is generated by
 protons through the synchrotron channel.}
\label{spectra}
\end{figure}

Thus within the suggested scenario, we expect a broad-band
SED consisting of three synchrotron peaks of different origin
located at the keV, MeV/GeV and TeV bands. Schematically,
such a SED is shown in Figure~\ref{spectra}. While the amplitudes of the
MeV and TeV synchrotron peaks are determined by the
total energy accelerated in the form of electrons and protons, respectively,
the intensity of the low energy synchrotron peak is determined by the
fraction of absorbed TeV (proton-synchrotron) gamma-rays.

\section{Application to \pkstitle}

In the case of the July 2006 flares of \pks{} the total energy of
the nonthermal radiation detected during the burst was about $E_{\rm
  tot}\approx L_{\gamma}\Delta t \approx 10^{51} \;\rm erg$.  According
to Equation~(\ref{eq:e_tot}), such an energy release can be produced by
an ensemble of clouds satisfying to the following condition:
\begin{equation}
\xi_{-1}M_{\rm c,25}\Gamma_{\rm j,1.5}^3\approx10\,.
\label{mass_condtion}
\end{equation}
This requirement can be easily fulfilled given the large mass of material,
that can be accelerated by the jet. Indeed, for the mass obtained
through Equation~(\ref{mcrce}), the above condition is reduced to:
\begin{equation}
L_{\rm j,46}>0.5 \frac{1}{\xi_{-1}M_{\rm BH,8}\Gamma_{\rm j,1.5}^2}\,.
\label{jet_condtion}
\end{equation}

The detected short variability of $\tau\sim 200$~s implies some
additional constraints on the system parameters. 
The causality condition constrains the size of the blob
$r_{\rm c}\approx\delta \tau c \approx \Gamma_{\rm j} \tau c$ 
(here we assume the size of the production region 
to be $2r_{\rm c}$ and $\d=2\Gamma_{\rm j}$). This allows one to relate the variability
time-scale and the peak luminosity through Equation~(\ref{eq:lgamma}):
\begin{equation}
 L_{\rm j,46} > 30\frac{M_{\rm BH,8}^2L_{\gamma,47}}{\tau_{2}^2\Gamma_{\rm j,1.5}^2 \xi_{-1}} \,.
\label{lj_condtion}
\end{equation}
In the specific case of \pks, this condition implies a lower limit
on the jet power, $L_{\rm j}>10^{47} M_{\rm BH,8}^2\Gamma_{\rm
  j,1.5}^{-2} \xi_{-1}^{-1} \mbox{ erg s}^{-1}$. Since  the Eddington 
luminosity has the following value  $L_{\rm Ed}\approx 1.4\times10^{46} M_{\rm BH,8} 
 \mbox{ erg s}^{-1}$, the jet should have a super Eddington luminosity, 
unless the bulk Lorentz factor is large, i.e $\Gamma_{\rm j}> 75 M_{\rm BH,8}^{1/2}\xi_{-1}^{-1/2} $. 


In the framework of JRGI scenario, the above conditions are quite
general and do not depend on the radiation mechanism.
Obviously, the available energy
has to be transferred to gamma-rays by a radiation mechanism with an
adequate cooling time. It was shown in Section~\ref{sec:proc} that, in general, both leptonic
(EIC) and hadronic (proton synchrotron) mechanisms can be quite
efficient in the frameworks of the JRGI scenario. Below we check 
the feasibility of these two mechanisms for the specific case of 
\pks; namely, we combine the energy and variability constraints 
with radiation mechanism specific limitations (e.g. cooling time).

\subsection{EIC model for \pkstitle}

\begin{figure}
\plotone{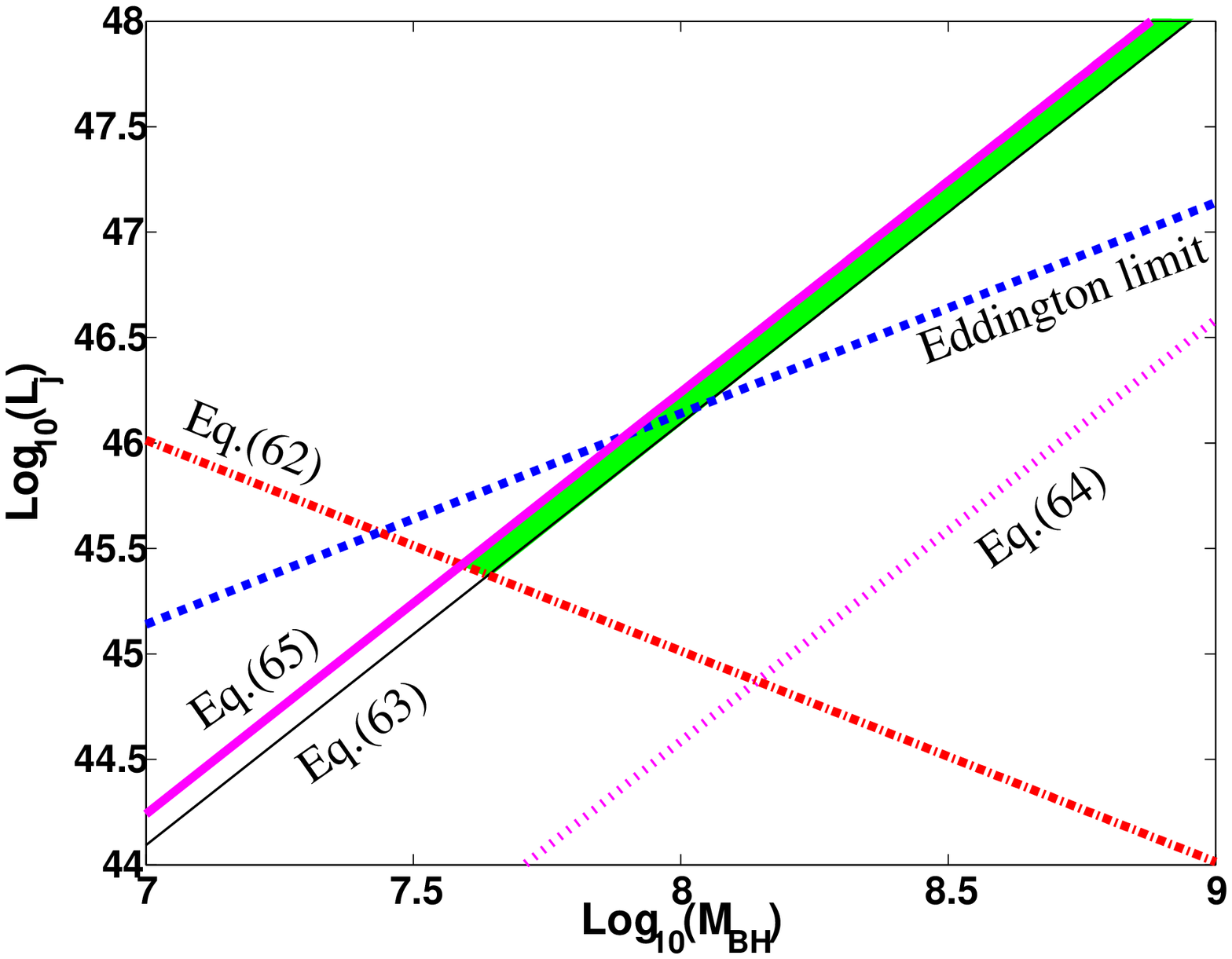}
\caption{Intensive VHE flares (in the case EIC) of 
$L_\gamma\approx10^{47}\,\rm erg\,s^{-1}$ and of duration $\tau\approx200$~s that
are energetically allowed in the framework of our JRGI scenario in systems 
characterized by a jet of luminosity, $L_{\rm j}$, powered by a black hole 
of mass $M_{\rm BH}$, are shown by the green filled region
(we assume $\Gamma_{\rm j}=90$ and $\xi_{-1}=1$).
The thin  solid line is the limit from Equation~(\ref{lj_condtion}); 
the thick  solid line is the limit from Equation~(\ref{lj_eic2}); 
the doted line is the limit from Equation~(\ref{lj_eic1}); 
the dot-dashed line is the limit from Equation~(\ref{jet_condtion}); and the dashed line 
is the Eddington limit. }
\label{ljmbh_eic}
\end{figure}

Given the short  cooling time in the case of EIC  (see Equation~(\ref{eq01})), 
the required variability can be achieved for a relatively low jet luminosity:
\begin{equation}
 L_{\rm j,46} > 0.007\frac{M_{\rm BH,8}^2 \Gamma_{\rm j,1.5}^{10/3}}{\tau_{2}^{4/3}\nu_{16}^{2/3}} \,.
\label{lj_eic1}
\end{equation}
On the other hand, the IC scattering should occur in the Thomson regime, 
otherwise the gamma-gamma opacity would significantly exceed  unity 
(see Equation~(\ref{kn_tau})). Thus, Equation~(\ref{z_eic}) allows the following
 upper limit on the jet luminosity:
 \begin{equation}
 L_{\rm j,46} \ll 0.4\frac{M_{\rm BH,8}^2 \Gamma_{\rm j,1.5}^{6}\nu_{16}^2}{E_{11}^{4}} \,,
\label{lj_eic2}
\end{equation}
which appears to be rather close to the lower limit given by
Equation~(\ref{lj_condtion}).  For the sake of clarity, we have combined
all the relevant limitations in Figure~\ref{ljmbh_eic}, where the
favorable parameters are indicated by the filled region.  It is
expected that the EIC mechanism can be responsible for nonthermal
emission with the required characteristics in the case of
sub-Eddington jets. On the other hand, the EIC model requires less
comfortable bulk Lorentz factor, which for the radiation production
region should exceed $\Gamma_{\rm j}>75 $. In the framework of the
JRGI scenario this implies that the star enters into the jet quite far
from the BH, $z>4\times 10^{17}M_{\rm BH,8}$~cm. Finally, we note that
the EIC model does not pose a strict requirement on the acceleration
process given the relatively small energy of the emitting
electron. 

\subsection{Proton synchrotron model for \pkstitle}

One of the most fundamental challenges for proton synchrotron 
models in AGN jets is the long cooling time. This is often interpreted  
as an indication of the extremely low efficiency of such models. 
In particular, \citet{s10psyn} has compared the synchrotron cooling time 
 to the expected adiabatic cooling time in relativistic jets. 
Based on this comparison, it was concluded that this mechanism was 
not feasible for nonthermal photon production in AGN jets. However, we 
note that this estimate depends strongly on the key assumption 
of the jet magnetization.  This parameter has been accepted to 
be quite low by \citet{s10psyn}, whilst in this paper, we rely 
on models of  magnetically driven  jets, i.e. the magnetization can be very high. Given the large magnetic field, proton 
synchrotron radiation from jets can be quite efficient in our scenario.

\begin{figure}
\plotone{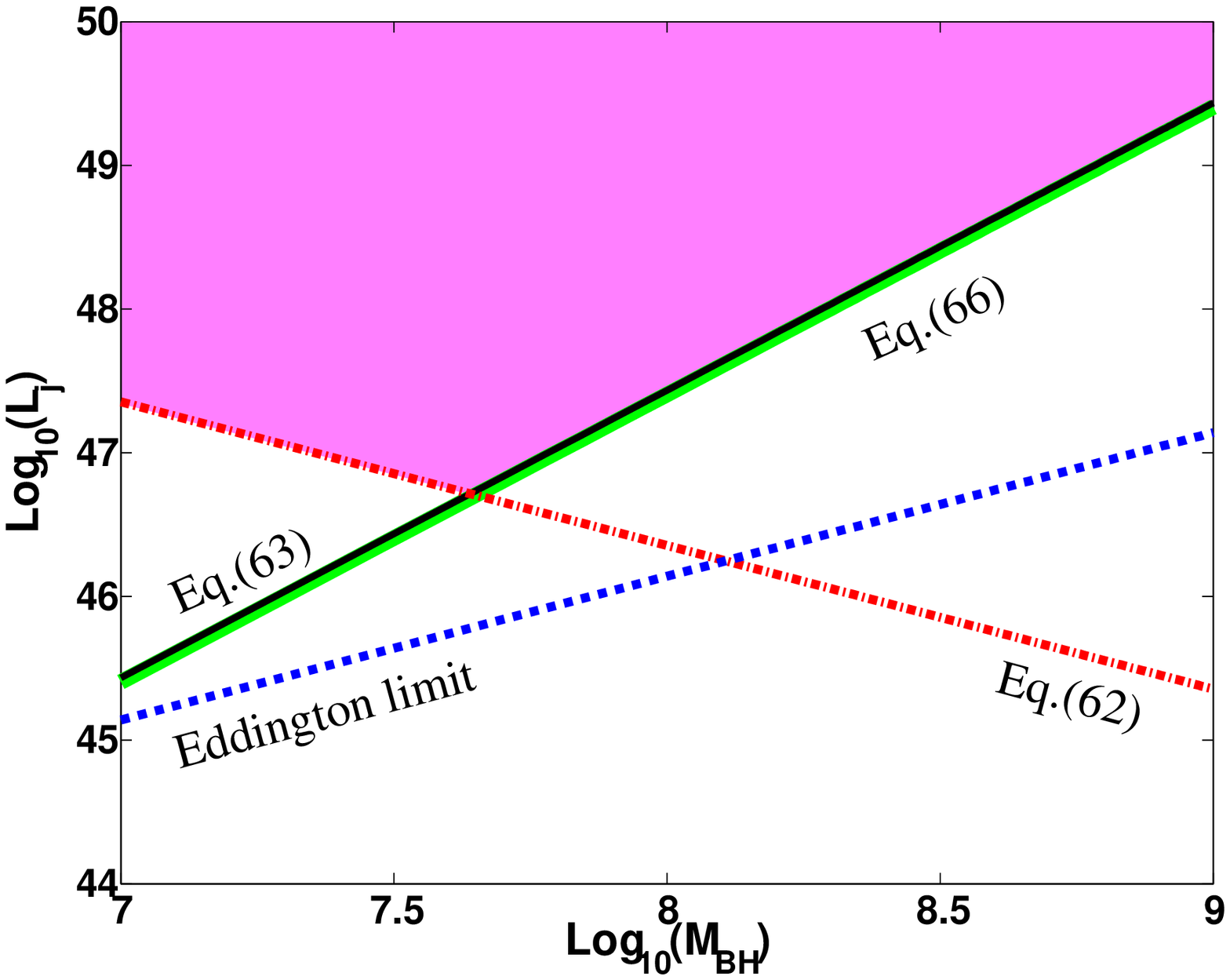}
\caption{Intensive VHE flares (in the case of proton synchrotron) 
of $L_\gamma\approx10^{47}\,\rm erg\,s^{-1}$ and of duration 
$\tau\approx200$~s that are energetically allowed in the framework 
of the JRGI scenario in systems characterized by a jet of luminosity, 
$L_{\rm j}$, powered by a black hole of mass $M_{\rm BH}$, 
which are shown by the pink filled region
(we assume $\Gamma_{\rm j}=20$ and $\xi_{-1}=1$).
The thin solid line is the limit from Equation~(\ref{lj_condtion}); 
the thick solid line is the limit from Equation~(\ref{lj_ps}); 
the dot-dashed line is the limit from Equation~(\ref{jet_condtion}); and the dashed line 
is the Eddington luminosity. }
\label{ljmbh_ps}
\end{figure}

The cooling time for proton synchrotron radiation, 
Equation~(\ref{syn_pr_cooling}),  recall for a very luminous jet
 \begin{equation}
 L_{\rm j,46} > 500\frac{M_{\rm BH,8}^{2} \Gamma_{\rm j,1.5}^{8/3}\eta_1^{2/3}}{\tau_2^{4/3}} \,.
\label{lj_ps}
\end{equation}
Note that, Equations~(\ref{jet_condtion}), (\ref{lj_condtion}) and (\ref{lj_ps}) 
 together require quite an extreme jet luminosity, but the requirement may be 
relaxed in the case of relatively small bulk Lorentz factors, 
i.e., $\Gamma_{\rm j}\approx 20 $. Although even in this case, the jet 
luminosity exceeds  the Eddington luminosity. The relevant 
parameter space is shown by the filled region in Figure~\ref{ljmbh_ps}. 
This corresponds to the case when the JRGI scenario is realized 
relatively close to the base of the jet, 
$z\approx 3\times10^{16}M_{\rm BH,8}$~cm. Finally, we note that 
proton synchrotron models require a very efficient particle 
acceleration with a rate close to the theoretical electrodynamic 
limit, i.e. $\eta \leq 10$.  The corresponding Doppler factor can 
be as high as $\delta\gtrsim 40$. One should expect a cutoff in the 
gamma-ray spectrum at around $\approx 1.5 \eta^{-1}_1$ TeV.  Thus, the 
observed VHE spectrum, extending up to 4 TeV, requires an acceleration 
parameter of $\eta \leq 10$. This implies that the proton synchrotron 
model can be realized only in extreme accelerators.

\section{Discussion}

The ultra-short TeV gamma-ray flares of blazars detected in the case
of \pks{} \citep{ah07pks} and Mkr 501 \citep{mkr501_magic} 
on 100~s timescales represent a serious challenge for current models 
of blazars. This challenge concerns the
origin and the sites of formation of these flares, the acceleration and
radiation mechanisms, the hydrodynamics of relativistic outflows,
amongst others. Since the upper limit on the size of production region, of $3 \times
10^{12} \tau_2 \ \rm cm$, is smaller by an order of magnitude than the
gravitational radius of a black hole of mass $10^{8} M_\odot$ (which
is required to power distant blazars), the only way to avoid the situation of invoking quite
uncomfortable upper limits on the mass of the central black hole (as
small as $10^7 M_\odot$), is to invoke the  Doppler
boosting. However, this can be realized only in the case of an external
origin of the processes which cause these ultra-short flares. If the
flares are initiated by disturbances originating from the central
black hole (e.g., due to internal shocks), the linear size (in the
observer's frame) of the flare production region cannot be smaller than
the gravitational radius of the black hole, independent of the
Doppler factor of the jet. The model suggested in this work readily
solves the problem of connecting the flares to the interactions of the red
giant stars with the powerful jets. Due to these interactions the red
giant loses a significant fraction its atmosphere. The cloud,
accelerated by the magnetically driven jet up to a Lorentz factor of
$\Gamma \sim 30$, likely separates into many small fragments. These
``blobs'' represent the ideal sites for the production of flares, provided that
a significant fraction of jet energy absorbed by the cloud is
converted (e.g. due to relativistic shock acceleration or magnetic
reconnection) to relativistic particles.

\begin{figure}
\plotone{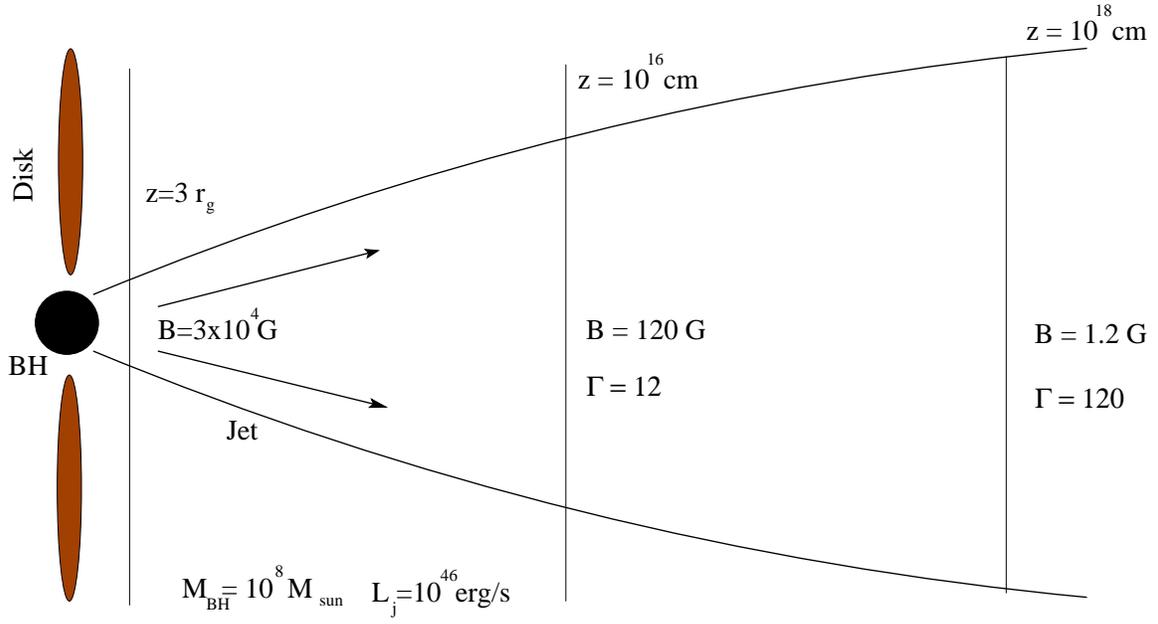}
\caption{Sketch of the jet together with characteristic magnetic field
  strength and bulk Lorentz factors at typical distances from the
  BH. The BH mass and the dimensionless parameter of rotation were
  assumed to be $M_{\rm BH}=10^8 M_{\odot}$ and $a\approx 1$,
  respectively. Initially, the jet is assumed to be magnetically
  dominated with the magnetization parameter $\sigma\gtrsim 100$.}
\label{magjet}
\end{figure}

The effective acceleration of particles is a necessary, but not a sufficient 
condition for the interpretation of the gamma-ray radiation of blazars.  
Any successful model of TeV blazars require adequate cooling 
times through gamma-radiation;  they should be comparable, or often 
even shorter compared to the characteristic timescales of other 
radiative and non-radiative processes. Generally, leptonic models 
of gamma-ray loud blazars, through the realization of SSC or external 
IC scenarios, do provide adequate radiation timescales, but at the expense 
of the assumption of a rather weak magnetic field, typically less than 1 G, 
which in powerful blazars ($L_{\rm j} \geq 10^{46}  \mbox{ erg s}^{-1}$), is well below 
the magnetic field in the jet as long as it's concerned with sub-parsec 
distances (see Figure~\ref{magjet}). This is a quite challenging requirement of the 
discussion of feasibility of, which is generally ignored in the 
literature. In the JRGI scenario suggested here the 
problem can be formally solved assuming that the magnetic field 
inside the blob is much smaller than in the jet. However, 
in the case of the SSC models,  this assumption still does not 
allow a relaxation of the second requirement of an extremely large 
jet Lorentz factor, $\Gamma \geq 1000$.
Although such Lorentz factors for the bulk motion  cannot
be excluded\footnote{ We should note
that it is rather difficult to reach such a high value of the bulk
Lorentz factor, e.g. due to the so called ``photon breeding
mechanism'' \citep{sp06}, which does not allow AGN jets with bulk
Lorentz factors exceeding 50.}, in particular at large, $\geq 1$pc distances 
from the BH  (see Figure~\ref{magjet}), in the
proposed JRGI model it hardly can work. At such distance the jet ram
pressure is not sufficient able to ablate the atmosphere of the star.

The requirements on the magnetic field and the jet Lorentz factor
are more relaxed in the external IC model. Nevertheless, one should 
note that within the JRGI scenario, the external Compton model
has some specific features. In order to  avoid severe gamma-gamma 
absorption, the Compton scattering should proceed in the Thomson regime.
This can be fulfilled if the radiation region is located at large distances, 
i.e. regions  still with quite a  large Lorenz factors for the jet,
$\Gamma_{\rm j} \sim 100$.

One of the main postulates of the JRGI scenario is the effective
star-jet interaction. This requires the location of the blobs that
emit gamma-rays to be at small distances from the BH, typically $z
\sim 10^{17}$~cm.  In the case of powerful jets, $L_{\rm j} \geq
10^{47} \mbox{ erg s}^{-1}$, this implies a very large magnetic field,
$B \sim 100$~G and a moderate Lorenz factor, $\Gamma_{\rm j} \sim
20$. Both parameters match nicely with the interpretation of the TeV
gamma-ray flares as a result of proton-synchrotron radiation by highly
magnetized blobs, formed and accelerated in jet-star
interactions. This model demands the acceleration of protons to
energies of $10^{19}$eV, and implies the acceleration of protons with
a rate close to the maximum (theoretically possible) rate of, $t_{\rm
  acc} \sim r_{\rm L}/c$. This is quite a robust requirement, which
however, can be provided, in principle, by certain acceleration
mechanisms.  Another challenge of the proposed scenario is related to
the power of the jet. Namely, the proton synchrotron model of TeV
gamma-rays can be effective, provided that: (i) the mass of BH does
not significantly exceed $M \sim 10^8 M_{\odot}$ and; (ii) the jet
power is not significantly below $10^{47} \mbox{ erg s}^{-1}$.  An
unambiguous implication of these two requirements (working in two
different directions) is that the jet should have a super-Eddington
luminosity. Although this could seem like quite a dramatic assumption,
we note the requirement of super-Eddington luminosities seems to be an
unavoidable, model-independent conclusion for GRBs and also likely for
powerful gamma-ray blazars \citep{g11texas}.

\subsection{Stellar density in the vicinity of a SMBH.}

An important question in the suggested scenario is the expected rate
of the flaring events, which is related to the number density of RGs at the
relevant jet scales. The jet region suitable for the production of the
powerful flares (similar to the burst detected from \pks), can
be defined as $z<1\rm pc$, and the corresponding side cross section of
the jet is $S_{\rm j}\approx z^2 \theta\sim 10^{33}\theta_{-1} z^2_{17}\rm cm^2$.
Thus, the number of flaring events per year can be estimated as  $\Upsilon \approx S_{\rm j} V_{\rm orb} n$.
Equation~(\ref{sv}) provides an estimate for  the density of  RGs required to produce $\Upsilon$ flaring
events per year:
\begin{equation}
n\sim10^6\Upsilon M_{\rm BH,8}^{-1/2}\theta^{-1}_{-1}z^{-3/2}_{17}\rm pc^{-3}\,.
\label{eq:density_RG}
\end{equation}
Unfortunately, there are no direct measurements of the stellar density
in the vicinity of BHs.  Thus, depending on the assumed
distribution law, the number of RGs in the vicinity of the BH may or
may not be sufficient. However, we note that studies of possible
stellar density profiles in the vicinity of the BH in AGNs (see
e.g. \citep{bkck82,mcd91}) show that densities similar to the one required
 ($\sim 10^6$~pc$^{-3}$) are rather feasible. Moreover, under the
influence of X-ray radiation close to BHs, normal stars can drastically
increase the rate of their stellar wind. Thus, wind-fed clouds can be
formed during the jet -- star interaction. This interaction can mimic
the interaction of RG atmosphere with the jet (Dorodnitsyn, private communications). Since,
the stellar density of normal stars is higher up to 2 orders of magnitude
 than the density
of RGs, this effect can significantly relax the requirement imposed by
Equation~(\ref{eq:density_RG}) on the stellar density in the vicinity of BHs.

\appendix
\section{Derivation of the cloud dynamics equation in the relativistic stage}
\label{der12}

Let us choose the orientation of coordinate systems such that in the
observer reference frame, $K$, the magnetic and electric field vectors
have the following components $ {\bf B} = (0, B, 0) $ and $ {\bf E} =
(E, 0,0) $, respectively. Since the conductivity of plasma is very
high, the following condition is held $ E = v_{\rm j}B / c $, where $ v_{\rm j} $
is jet velocity.  The cloud velocity in the system $K$ is $v_{\rm c}$,
and $K'$ is the cloud's momentarily comoving reference frame.
Quantities pertaining to the system $K'$ are marked by prime.  In the
system, $K'$ electromagnetic field strengths are
\begin{gather}
E'=\Gamma_{\rm c}\left(E-\frac{v_{\rm c}}{c}B\right)=\frac1c\Gamma_{\rm c}(v_{\rm j}-v_{\rm c})B\,,\\
B'=\Gamma_{\rm c}\left(B-\frac{v_{\rm c}}{c}E\right)=\Gamma_{\rm c}\left(1-\frac{v_{\rm j}v_{\rm c}}{c^2}\right)B\,.
\end{gather}
As it should be, the following ratio holds; $ E'/ B' = v'_{\rm j} / c $,
where $ v'_{\rm j} $ is the jet speed relative to the cloud. Energy flux
density in system $ K'$ and in the laboratory system are related
by:
\begin{equation}
q'=\frac{c}{4\pi}E'B'=\frac1c\Gamma_{\rm c}^2(v_{\rm j}-v_{\rm c})(1-v_{\rm j}v_{\rm c}/c^2)\,q\,.
\end{equation}
Jet ram pressure in the system $ K'$ is equal to $ q' / c $, thus
during a differentially small time interval, $ dt'$, the cloud momentum
increases by a value of $ dP'_{\rm c} = (q' / c) \, \pi r_{\rm c}^2 \, dt'$, and
the energy increment is a second order value of $dt'$, i.e $ dE'_{\rm c} = 0
$. In the observer system, one has $ dE_{\rm c} = M_{\rm c} c^2 d \Gamma_{\rm c} =
\Gamma_{\rm c} v_{\rm c} dP'_{\rm c} $, and $ dt = \Gamma_{\rm c} dt'$. Thus, the equation of
motion may be expressed as follows:
\begin{equation}
\frac{d\Gamma_{\rm c}}{dt}=\frac{\pi r_{\rm c}^2 v_{\rm c}}{M_{\rm c} c^3}\,q'=
\frac{\pi r_{\rm c}^2}{M_{\rm c} c^4}\,v_{\rm c}\Gamma_{\rm c}^2 (v_{\rm j}-v_{\rm c})(1-v_{\rm j} v_{\rm c}/c^2)\,q\,.
\end{equation}
Denoting $ q = L_{\rm j} / \pi \omega^2 $ and assuming $ \Gamma_{\rm j} \gg 1 $
and $ \Gamma_{\rm c} \gg 1 $, one obtains the equation of motion in the form
of Equation~(\ref{dgdt})
\begin{equation}
\frac{d\Gamma_{\rm c}}{dt}=\left(\frac{1}{\Gamma_{\rm c}^2}-\frac{\Gamma_{\rm c}^2}{\Gamma^4_{\rm j}}\right)\frac
{L_{\rm j} r_{\rm c}^2}{4\omega^2 c^2 M_{\rm c}}\,\rm,
\label{dgdt_copy}
\end{equation}

\section{The variation of $\d$ in the case of blob chaotic motion}
\label{aboost}
In the suggested scenario, non-thermal particle acceleration is
triggered by RG material blobs caught up in the jet. In such a case it
is natural to assume that the acceleration sites are closely related
to the blobs.  Although the particle propagation and isotropisation
processes may be very complicated, the highest energy particles lose
their energy very close to the acceleration site. Thus, in what
follows we assume that the VHE emission is related to the blobs and the
radiation boosting factor is determined by the blob velocity.  Since
the blob velocity can be changed relatively easy, sudden changes of
the Doppler factor are rather feasible.  Hence we have discuss
the related changes of the Doppler factor and {\it correction function
  $F_{\rm e}$}, i.e. the quantities describing the non-thermal
emission intensity, as seen by observer.

\subsection{A}

Let us assume that at the moment $t_0$, a blob moves relativistically
in the direction of the observer with velocity $V=\beta c$, thus, the
corresponding Lorentz factor is $\Gamma=1/\sqrt{1-V^2}\gg1$.  Let us
consider two reference frames: the observer coordinate system $K$, and
the blob comoving system $K'$, where the blob is at
rest at $t_0$. The $z$ axises are selected to be parallel
to the systems' relative velocity, $\mathbf V$. Let us consider two
cases for a change of the blob velocity.

In the $K'$ reference frame, the blob gains velocity $\mathbf v'$, such as: 
\begin{equation}
v'_x=c\beta'\sin\phi\,,\quad v'_y=0\,,\quad v'_z=c\beta'\cos\phi\,.
\label{bb1}
\end{equation}
Here  $ \phi $ is the angle between the velocity $\mathbf v'$ and $z'$ axis; 
$\beta' \equiv v'/c$; and the observer detects the emission radiated along 
$ z $ axis. In the laboratory frame, the blob velocity components have the following form:
\begin{equation}
v_x=\frac{c}{\Gamma}\,\frac{\beta'\sin\phi}{1+\beta\beta'\cos\phi}\,,\quad
v_y=0\,,\quad
v_z=c\,\frac{\beta+\beta'\cos\phi}{1+\beta\beta'\cos\phi}\,.
\label{bb2}
\end{equation}
Therefore, the blob Lorentz factor is
$\tilde\Gamma=1/\sqrt{1-(v_x^2+v_y^2+v_z^2)/c^2}$. Given 
that the system is in the
ultrarelativitic regime (in the Appendixes we will assume $\Gamma\gg 1$), 
one obtains that the Doppler factor is:
\begin{equation}
\d=\left[\tilde\Gamma(1-v_z/c)\right]^{-1}=
\frac{2\Gamma\sqrt{1-\beta'^2}}{1-\beta'\cos\phi}\,.
\label{bb3}
\end{equation}
In the laboratory frame the velocity deflection angle is
$\d\theta=\beta'\sin\phi/\Gamma\ll1$.
Thus, the Lorentz factor of the radial (i.e. along the $ z $ axis) motion
$\Gamma_z$ can be expressed as
\begin{equation}
\Gamma_z\equiv\frac1{\sqrt{1-v_z^2/c^2}}=\Gamma\sqrt{\frac{1+\beta'\cos\phi}{
1-\beta'\cos\phi}}\,.
\label{bb4}
\end{equation}
The radiation intensity is defined by the {\it correction function},
if the time-scale of the blob velocity variation is longer than the typical
particle acceleration time. In the case of  $\Gamma_{\rm j}\gg \Gamma_{\rm c} \gg 1$, one obtains
\begin{equation}
r(\phi)=\frac{\d^4}{\Gamma_z^2}=\frac{16\Gamma^2(1-\beta'^2)^2}
{(1+\beta'\cos\phi)(1-\beta'\cos\phi)^3}\, ,
\label{bb5}
\end{equation}
Otherwise (i.e. if the particle cooling time is long as compared to
the time-scale of the velocity change), the apparent luminosity is
proportional to the standard factor $\d^4$. Thus, one obtains 
\begin{equation}
r(\phi)=\d^4=\frac{16\Gamma^4(1-\beta'^2)^2}
{(1-\beta'\cos\phi)^4}\, .
\label{bb5b}
\end{equation}
As it can be seen from the comparison of Equations~(\ref{bb5}) and
(\ref{bb5b}), the function $r$ has a weak dependence on the velocity change. At the moment
$t_0$, this function has a value of  $r_0\equiv
r(\phi)\big|_{\beta'=0}=16\Gamma^2$. The change of the velocity may
lead both to an increase and decrease of the function $r(\phi)$: if $\phi=0$ and
$\beta'\ge1/3$, the correction function value is $r(0)\ge 2r_0$; and if
$\phi=\pi/2$ and $\beta'\ge 0.54$, one has $r(\pi/2)\le r_0/2$.




\subsection{B}

Another potentially important situation occurs when a blob moves along
a helical trajectory, e.g. along the dominant magnetic field line. In this
case, the additional velocity component is oriented perpendicularly to
the averaged velocity, i.e. its components may be expressed as the
follows:
\begin{equation}
  v'_x=c\beta'\cos\phi\,,\quad v'_y=c\beta'\sin\phi\,,\quad v'_z=0\,.
\label{bb9}
\end{equation}
Obviously, if an observer is located in the direction of the $z'$
axis, no flux variability  is detected. On the other hand, if the observer is located
slightly off axis, a periodic increase of the flux
level is seen. Let us assume that the observer is located in the direction
of the blob motion for $\phi=0$. In this case, the Doppler factor has
the following form:
\begin{equation}
\d=\frac{2\Gamma\sqrt{1-\beta'^2}}{1-\beta'^2(2\cos\phi-1)}\,.
\label{bb10}
\end{equation}
The radiation intensity, then, is proportional to the following factor:
\begin{equation}
r=\frac{\d^4}{\Gamma^2}=\frac{16\Gamma^2(1-\beta'^2)^2}{
\big(1-\beta'^2(2\cos\phi-1)\big)^4}\,.
\label{bb11}
\end{equation}

\begin{figure}
\begin{center}
\includegraphics[width=0.7\textwidth,angle=-90]{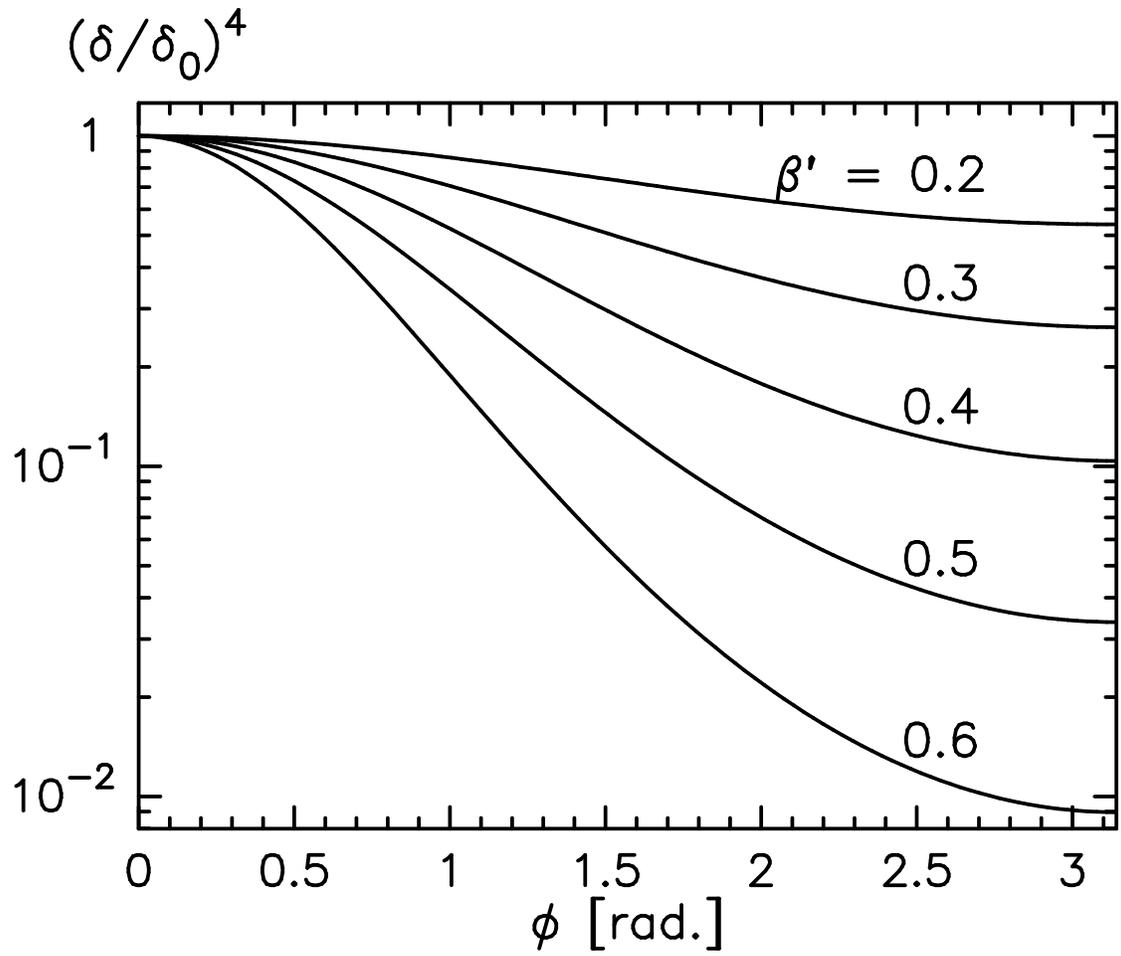}
\caption{Dependence of the observed emission intensity, as defined by Equation~(\ref{bb11}), on the position of the
blob. The intensities are normalized to the maximum values (i.e the
value at $ \phi = 0 $).
The values of the parameter $ \beta'$ are shown in the figure. }
\label{delta_b}
\end{center}
\end{figure}

Obviously, the maximum value of $ r $ is achieved when $ \phi = 0 $, i.e. 
when the radiation is emitted towards the observer. If the precession  
velocity fulfills the following condition
\begin{equation}
\tilde\beta'>\left(\frac{2^{1/4}-1}{2^{1/4}+1-2\cos\phi}\right)^{1/2}\,,
\label{bb12}
\end{equation}
then at the corresponding moment, the emission intensity can differ by
a factor of 2 as compared to the maximum. For example, if $ \phi = \pi/4
$ then the required velocity value is $ \tilde \beta' \approx0.5 $. In
Figure~\ref{delta_b} we show the $\phi$-angle dependence of the
normalized intensity for several different values of $ \beta'$.

Another important issue is the time modulation of the emission, as seen
in the observer reference frame. Let us assume that the azimuthal angle
has the following time dependence; $\phi = \omega t'$, where $\omega$
is a constant; and $t'$ is time in the system $K'$. Then a quantity $
T_0 = 2 \pi / \omega $ corresponds to the precession period in the
system $K'$.  Obviously, in the observer frame the emission intensity
should have a different time dependence than the one that can be
obtained by the substitution of $\phi = \omega t'$ into the function shown
in Figure~\ref{delta_b}. Indeed, the radiation emitted at time 
$ t_r $, by the source is detected by an observer located at $ \b r = r
\b n $ at time $ t $, where
\begin{equation}\label{bbx1}
t=t_r+|\b r-\b r(t_r)|/c\approx t_r+r/c-(\b n\b r(t_r))/c\,,
\end{equation}
where $ \b r (t_r) $ is the emitter position at time $ t_r $.
Since the constant term $ r / c $ can be
neglected, Equation~(\ref{bbx1}) can be reduced to the following form:
\begin{equation}\label{bbx2}
t=t_r-(\b n\b r(t_r))/c\,.
\end{equation}

\begin{figure}
\begin{center}
\includegraphics[width=0.7\textwidth,angle=-90]{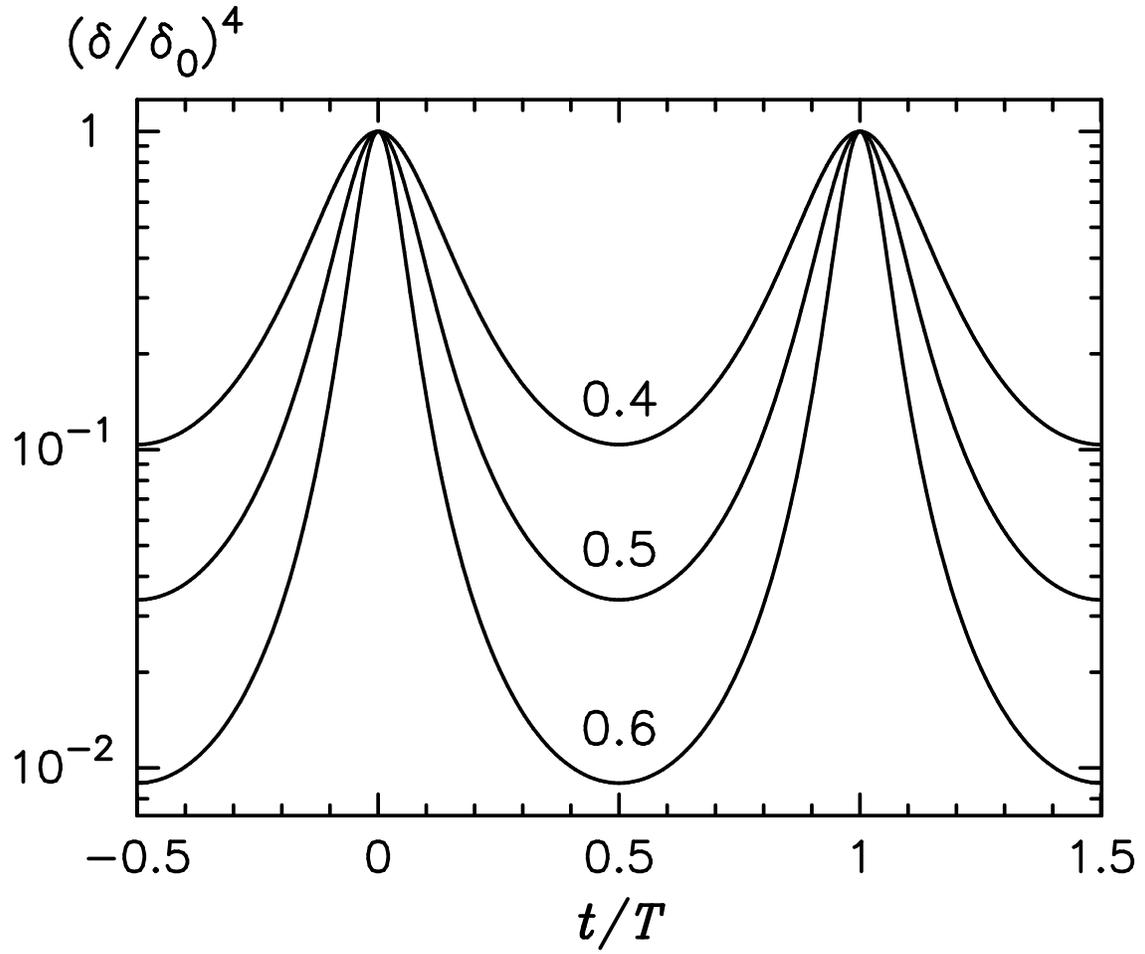}
\caption{ Dependence of the observed emission intensity (i.e. Equation~(\ref{bb11})) as a function of time, as obtained in Equation~(\ref{bbx3}), for
three different values of the $ \beta'$ parameter: $\beta'=0.4$,
$0.5$ and $0.6$ (in the case B).}
\label{delta_t}
\end{center}
\end{figure}

\begin{figure}
\begin{center}
\includegraphics[width=0.7\textwidth,angle=-90]{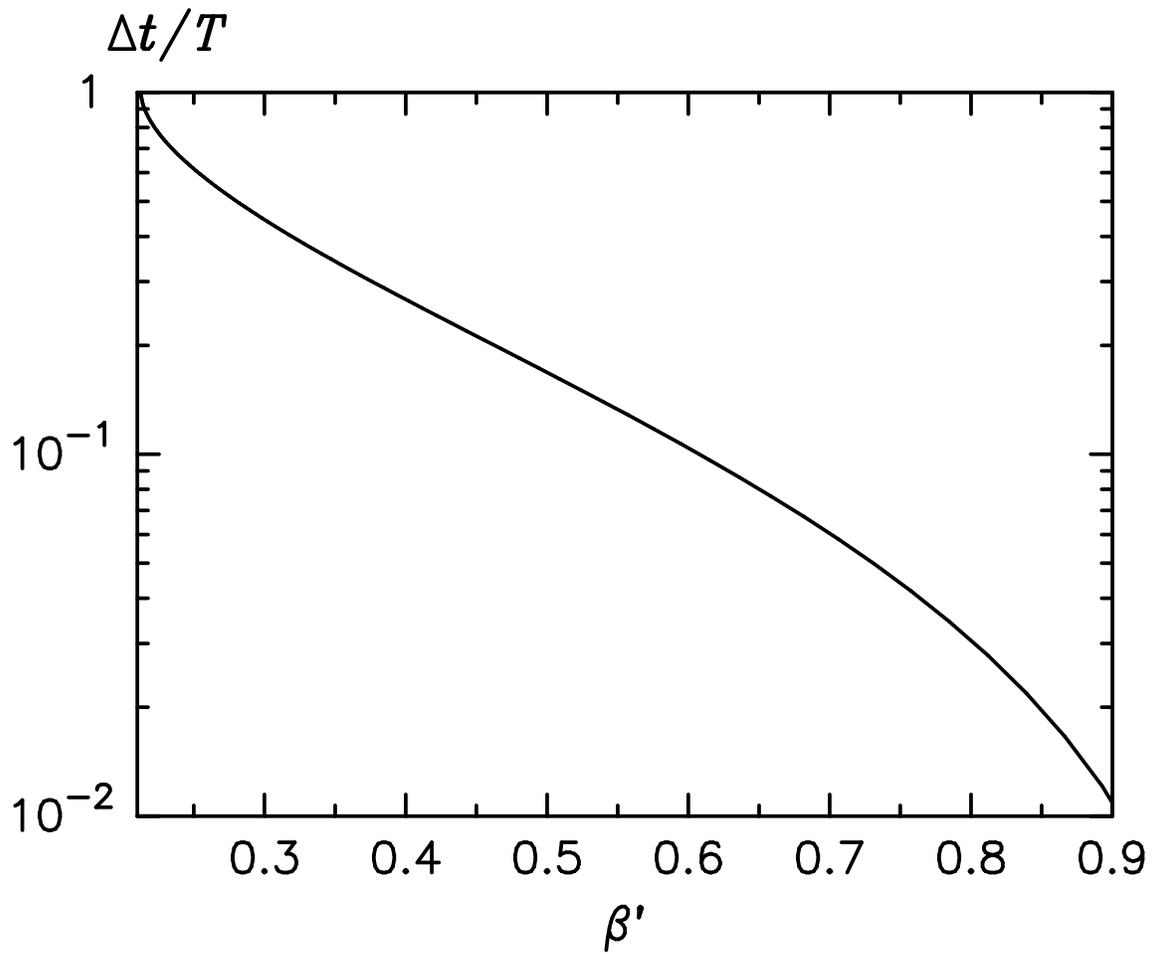}
\caption{The width of the intensity peaks, shown in Figure~\ref{delta_t}, as function of
$ \beta'$ (for details, see Appendix \ref{aboost}, case B).}
\label{t2}
\end{center}
\end{figure}

For $K$-frame quantities,  the azimuthal angle can be represented  as $ \phi = \omega t_r /
\Gamma $. This relationship allows us to
express $ t $ through $\phi$:
\begin{equation}\label{bbx3}
t=\frac{T_0(1+\beta'^2)}{4\pi\Gamma}\left(\phi-\frac{2\beta'^2}{1+\beta'^2}
\sin\phi\right)\,.
\end{equation}
By increasing $ \phi $ by $ 2 \pi $, the time, $ t $, changes by an
amount equivalent to:
\begin{equation}\label{bbx4}
T=\frac{T_0(1+\beta'^2)}{2\Gamma}\,,
\end{equation}
which is the period of the observed emission. The dependence of the
observed intensity, as a function of the observer time, is shown in Figure~\ref{delta_t}. As it
can be seen from this figure, the distribution width decreases with
an increase of $ \beta'$. The emission ``half-decay'' interval, $ \Delta t/ 2 $,
is shown in Figure~\ref{t2}  as a function of $\beta'$.

\section*{Acknowledgments}
The work of S.V.Bogovalov have been supported by the Federal Targeted Program
``The Scientific and Pedagogical Personnel of the Innovative Russia'' in 2009-2013 (the state
contract N 536 on May 17, 2010)





\begin{thebibliography}{47}
\expandafter\ifx\csname natexlab\endcsname\relax\def\natexlab#1{#1}\fi

\bibitem[{{Achterberg} {et~al.}(2001){Achterberg}, {Gallant}, {Kirk}, \&
  {Guthmann}}]{agk01}
{Achterberg}, A., {Gallant}, Y.~A., {Kirk}, J.~G., \& {Guthmann}, A.~W. 2001,
  \mnras, 328, 393

\bibitem[{{Aharonian} {et~al.}(2007){Aharonian}, {Akhperjanian}, {Bazer-Bachi},
  {Behera}, {Beilicke}, {Benbow}, {Berge}, {Bernl{\"o}hr}, {Boisson}, {Bolz},
  {Borrel}, {Boutelier}, {Braun}, {Brion}, {Brown}, {B{\"u}hler},
  {B{\"u}sching}, {Bulik}, {Carrigan}, {Chadwick}, {Clapson}, {Chounet},
  {Coignet}, {Cornils}, {Costamante}, {Degrange}, {Dickinson},
  {Djannati-Ata{\"i}}, {Domainko}, {Drury}, {Dubus}, {Dyks}, {Egberts},
  {Emmanoulopoulos}, {Espigat}, {Farnier}, {Feinstein}, {Fiasson},
  {F{\"o}rster}, {Fontaine}, {Funk}, {Funk}, {F{\"u}{\ss}ling}, {Gallant},
  {Giebels}, {Glicenstein}, {Gl{\"u}ck}, {Goret}, {Hadjichristidis}, {Hauser},
  {Hauser}, {Heinzelmann}, {Henri}, {Hermann}, {Hinton}, {Hoffmann}, {Hofmann},
  {Holleran}, {Hoppe}, {Horns}, {Jacholkowska}, {de Jager}, {Kendziorra},
  {Kerschhaggl}, {Kh{\'e}lifi}, {Komin}, {Kosack}, {Lamanna}, {Latham}, {Le
  Gallou}, {Lemi{\`e}re}, {Lemoine-Goumard}, {Lenain}, {Lohse}, {Martin},
  {Martineau-Huynh}, {Marcowith}, {Masterson}, {Maurin}, {McComb}, {Moderski},
  {Moulin}, {de Naurois}, {Nedbal}, {Nolan}, {Olive}, {Orford}, {Osborne},
  {Ostrowski}, {Panter}, {Pedaletti}, {Pelletier}, {Petrucci}, {Pita},
  {P{\"u}hlhofer}, {Punch}, {Ranchon}, {Raubenheimer}, {Raue}, {Rayner},
  {Renaud}, {Ripken}, {Rob}, {Rolland}, {Rosier-Lees}, {Rowell}, {Rudak},
  {Ruppel}, {Sahakian}, {Santangelo}, {Saug{\'e}}, {Schlenker}, {Schlickeiser},
  {Schr{\"o}der}, {Schwanke}, {Schwarzburg}, {Schwemmer}, {Shalchi}, {Sol},
  {Spangler}, {Stawarz}, {Steenkamp}, {Stegmann}, {Superina}, {Tam},
  {Tavernet}, {Terrier}, {van Eldik}, {Vasileiadis}, {Venter}, {Vialle},
  {Vincent}, {Vivier}, {V{\"o}lk}, {Volpe}, {Wagner}, {Ward}, \&
  {Zdziarski}}]{ah07pks}
{Aharonian}, F., {et~al.} 2007, \apjl, 664, L71

\bibitem[{{Aharonian} {et~al.}(2009{\natexlab{a}}){Aharonian}, {Akhperjanian},
  {Anton}, {Barres de Almeida}, {Bazer-Bachi}, {Becherini}, {Behera}, {Benbow},
  {Bernl{\"o}hr}, {Boisson}, {Bochow}, {Borrel}, {Brion}, {Brucker}, {Brun},
  {B{\"u}hler}, {Bulik}, {B{\"u}sching}, {Boutelier}, {Chadwick},
  {Charbonnier}, {Chaves}, {Cheesebrough}, {Chounet}, {Clapson}, {Coignet},
  {Costamante}, {Dalton}, {Daniel}, {Davids}, {Degrange}, {Deil}, {Dickinson},
  {Djannati-Ata{\"i}}, {Domainko}, {O'C.~Drury}, {Dubois}, {Dubus}, {Dyks},
  {Dyrda}, {Egberts}, {Emmanoulopoulos}, {Espigat}, {Farnier}, {Feinstein},
  {Fiasson}, {F{\"o}rster}, {Fontaine}, {F{\"u}{\ss}ling}, {Gabici}, {Gallant},
  {G{\'e}rard}, {Giebels}, {Glicenstein}, {Gl{\"u}ck}, {Goret}, {G{\"o}hring},
  {Hauser}, {Hauser}, {Heinz}, {Heinzelmann}, {Henri}, {Hermann}, {Hinton},
  {Hoffmann}, {Hofmann}, {Holleran}, {Hoppe}, {Horns}, {Jacholkowska}, {de
  Jager}, {Jahn}, {Jung}, {Katarzy{\'n}ski}, {Katz}, {Kaufmann}, {Kendziorra},
  {Kerschhaggl}, {Khangulyan}, {Kh{\'e}lifi}, {Keogh}, {Klu{\'z}niak},
  {Kneiske}, {Komin}, {Kosack}, {Lamanna}, {Lenain}, {Lohse}, {Marandon},
  {Martin}, {Martineau-Huynh}, {Marcowith}, {Maurin}, {McComb}, {Medina},
  {Moderski}, {Monard}, {Moulin}, {Naumann-Godo}, {de Naurois}, {Nedbal},
  {Nekrassov}, {Niemiec}, {Nolan}, {Ohm}, {Olive}, {de O{\~n}a Wilhelmi},
  {Orford}, {Ostrowski}, {Panter}, {Paz Arribas}, {Pedaletti}, {Pelletier},
  {Petrucci}, {Pita}, {P{\"u}hlhofer}, {Punch}, {Quirrenbach}, {Raubenheimer},
  {Raue}, {Rayner}, {Renaud}, {Rieger}, {Ripken}, {Rob}, {Rosier-Lees},
  {Rowell}, {Rudak}, {Rulten}, {Ruppel}, {Sahakian}, {Santangelo},
  {Schlickeiser}, {Sch{\"o}ck}, {Schr{\"o}der}, {Schwanke}, {Schwarzburg},
  {Schwemmer}, {Shalchi}, {Sikora}, {Skilton}, {Sol}, {Spangler}, {Stawarz},
  {Steenkamp}, {Stegmann}, {Superina}, {Szostek}, {Tam}, {Tavernet}, {Terrier},
  {Tibolla}, {Tluczykont}, {van Eldik}, {Vasileiadis}, {Venter}, {Venter},
  {Vialle}, {Vincent}, {Vivier}, {V{\"o}lk}, {Volpe}, {Wagner}, {Ward},
  {Zdziarski}, \& {Zech}}]{ah_pks_09}
---. 2009{\natexlab{a}}, {A\&A}, 502, 749

\bibitem[{{Aharonian} {et~al.}(2009{\natexlab{b}}){Aharonian}, {Akhperjanian},
  {Anton}, {Barres de Almeida}, {Bazer-Bachi}, {Becherini}, {Behera},
  {Bernl{\"o}hr}, {Boisson}, {Bochow}, \& {et al.}}]{a_pks_low_09}
---. 2009{\natexlab{b}}, \apjl, 696, L150

\bibitem[{{Aharonian}(2000)}]{ah00}
{Aharonian}, F.~A. 2000, New Astronomy, 5, 377

\bibitem[{{Aharonian} {et~al.}(2002){Aharonian}, {Belyanin}, {Derishev},
  {Kocharovsky}, \& {Kocharovsky}}]{abd02}
{Aharonian}, F.~A., {Belyanin}, A.~A., {Derishev}, E.~V., {Kocharovsky}, V.~V.,
  \& {Kocharovsky}, V.~V. 2002, \prd, 66, 023005

\bibitem[{{Aharonian} {et~al.}(2008){Aharonian}, {Khangulyan}, \&
  {Costamante}}]{akc08}
{Aharonian}, F.~A., {Khangulyan}, D., \& {Costamante}, L. 2008, \mnras, 387,
  1206

\bibitem[{Albert {et~al.}(2007)Albert, Aliu, Anderhub, Antoranz, Armada,
  Baixeras, Barrio, Bartko, Bastieri, Becker, Bednarek, Berger, Bigongiari,
  Biland, Bock, Bordas, Bosch-Ramon, Bretz, Britvitch, Camara, Carmona,
  Chilingarian, Coarasa, Commichau, Contreras, Cortina, Costado, Curtef,
  Danielyan, Dazzi, Angelis, Delgado, {de los Reyes}, Lotto,
  Domingo-Santamar{\'i}a, Dorner, Doro, Errando, Fagiolini, Ferenc,
  Fern{\'a}ndez, Firpo, Flix, Fonseca, Font, Fuchs, Galante,
  Garc{\'i}a-L{\'o}pez, Garczarczyk, Gaug, Giller, Goebel, Hakobyan, Hayashida,
  Hengstebeck, Herrero, H{\"o}hne, Hose, Hrupec, Hsu, Jacon, Jogler, Kosyra,
  Kranich, Kritzer, Laille, Lindfors, Lombardi, Longo, L{\'o}pez, L{\'o}pez,
  Lorenz, Majumdar, Maneva, Mannheim, Mansutti, Mariotti, Mart{\'i}nez, Mazin,
  Merck, Meucci, Meyer, Miranda, Mirzoyan, Mizobuchi, Moralejo, Nieto, Nilsson,
  Ninkovic, O{\~n}a-Wilhelmi, Otte, Oya, Paneque, Panniello, Paoletti, Paredes,
  Pasanen, Pascoli, Pauss, Pegna, Persic, Peruzzo, Piccioli, Prandini,
  Puchades, Raymers, Rhode, Rib{\'o}, Rico, Rissi, Robert, R{\"u}gamer,
  Saggion, Saito, S{\'a}nchez, Sartori, Scalzotto, Scapin, Schmitt, Schweizer,
  Shayduk, Shinozaki, Shore, Sidro, Sillanp{\"a}{\"a}, Sobczynska, Stamerra,
  Stark, Takalo, Tavecchio, Temnikov, Tescaro, Teshima, Torres, Turini, Vankov,
  Vitale, Wagner, Wibig, Wittek, Zandanel, Zanin, \& Zapatero}]{mkr501_magic}
Albert, J., {et~al.} 2007, The Astrophysical Journal, 669, 862

\bibitem[{{Araudo} {et~al.}(2010){Araudo}, {Bosch-Ramon}, \& {Romero}}]{abr10}
{Araudo}, A.~T., {Bosch-Ramon}, V., \& {Romero}, G.~E. 2010, A\&A, 522, A97+

\bibitem[{{Barkov} {et~al.}(2010){Barkov}, {Aharonian}, \&
  {Bosch-Ramon}}]{bab10}
{Barkov}, M.~V., {Aharonian}, F.~A., \& {Bosch-Ramon}, V. 2010, ArXiv e-prints

\bibitem[{{Barkov} \& {Komissarov}(2008)}]{bk08}
{Barkov}, M.~V., \& {Komissarov}, S.~S. 2008, International Journal of Modern
  Physics D, 17, 1669

\bibitem[{{Begelman} {et~al.}(2008){Begelman}, {Fabian}, \& {Rees}}]{bfr08}
{Begelman}, M.~C., {Fabian}, A.~C., \& {Rees}, M.~J. 2008, \mnras, 384, L19

\bibitem[{Beskin(2010)}]{bes10}
Beskin, V.~S. 2010, Physics-Uspehi, 180, 1241

\bibitem[{{Beskin} \& {Nokhrina}(2006)}]{bn06}
{Beskin}, V.~S., \& {Nokhrina}, E.~E. 2006, \mnras, 367, 375

\bibitem[{{Bisnovatyi-Kogan} {et~al.}(1982){Bisnovatyi-Kogan}, {Churaev}, \&
  {Kolosov}}]{bkck82}
{Bisnovatyi-Kogan}, G.~S., {Churaev}, R.~S., \& {Kolosov}, B.~I. 1982, {A\&Ap},
  113, 179

\bibitem[{{Blandford} \& {Znajek}(1977)}]{BZ77}
{Blandford}, R.~D., \& {Znajek}, R.~L. 1977, \mnras, 179, 433

\bibitem[{{Derishev}(2009)}]{der09}
{Derishev}, E.~V. 2009, International Journal of Modern Physics D, 18, 1523

\bibitem[{{Dermer} {et~al.}(2009){Dermer}, {Finke}, {Krug}, \&
  {B{\"o}ttcher}}]{dfkb09}
{Dermer}, C.~D., {Finke}, J.~D., {Krug}, H., \& {B{\"o}ttcher}, M. 2009, \apj,
  692, 32

\bibitem[{{Ghisellini}(2011)}]{g11texas}
{Ghisellini}, G. 2011, ArXiv e-prints

\bibitem[{{Giannios} {et~al.}(2009){Giannios}, {Uzdensky}, \&
  {Begelman}}]{gub09}
{Giannios}, D., {Uzdensky}, D.~A., \& {Begelman}, M.~C. 2009, \mnras, 395, L29

\bibitem[{{Gregori} {et~al.}(2000){Gregori}, {Miniati}, {Ryu}, \&
  {Jones}}]{gmr00}
{Gregori}, G., {Miniati}, F., {Ryu}, D., \& {Jones}, T.~W. 2000, \apj, 543, 775

\bibitem[{{Haswell} {et~al.}(1992){Haswell}, {Tajima}, \& {Sakai}}]{hts92}
{Haswell}, C.~A., {Tajima}, T., \& {Sakai}, J. 1992, \apj, 401, 495

\bibitem[{{Jorstad} {et~al.}(2001){Jorstad}, {Marscher}, {Mattox}, {Wehrle},
  {Bloom}, \& {Yurchenko}}]{J01}
{Jorstad}, S.~G., {Marscher}, A.~P., {Mattox}, J.~R., {Wehrle}, A.~E., {Bloom},
  S.~D., \& {Yurchenko}, A.~V. 2001, \apjs, 134, 181

\bibitem[{{Jorstad} {et~al.}(2005){Jorstad}, {Marscher}, {Lister}, {Stirling},
  {Cawthorne}, {Gear}, {G{\'o}mez}, {Stevens}, {Smith}, {Forster}, \&
  {Robson}}]{J05}
{Jorstad}, S.~G., {et~al.} 2005, \aj, 130, 1418

\bibitem[{{Komissarov} {et~al.}(2007{\natexlab{a}}){Komissarov}, {Barkov}, \&
  {Lyutikov}}]{kbl07}
{Komissarov}, S.~S., {Barkov}, M., \& {Lyutikov}, M. 2007{\natexlab{a}},
  \mnras, 374, 415

\bibitem[{{Komissarov} {et~al.}(2007{\natexlab{b}}){Komissarov}, {Barkov},
  {Vlahakis}, \& {K{\"o}nigl}}]{kbvk07}
{Komissarov}, S.~S., {Barkov}, M.~V., {Vlahakis}, N., \& {K{\"o}nigl}, A.
  2007{\natexlab{b}}, \mnras, 380, 51

\bibitem[{{Komissarov} {et~al.}(2010){Komissarov}, {Vlahakis}, \&
  {K{\"o}nigl}}]{kvk10}
{Komissarov}, S.~S., {Vlahakis}, N., \& {K{\"o}nigl}, A. 2010, \mnras, 407, 17

\bibitem[{{Komissarov} {et~al.}(2009){Komissarov}, {Vlahakis}, {K{\"o}nigl}, \&
  {Barkov}}]{kvkb09}
{Komissarov}, S.~S., {Vlahakis}, N., {K{\"o}nigl}, A., \& {Barkov}, M.~V. 2009,
  \mnras, 394, 1182

\bibitem[{{Levinson} \& {Bromberg}(2008)}]{lb08}
{Levinson}, A., \& {Bromberg}, O. 2008, International Journal of Modern Physics
  D, 17, 1603

\bibitem[{{Lobanov}(1998)}]{lob98}
{Lobanov}, A.~P. 1998, \aap, 330, 79

\bibitem[{{Lovelace}(1976)}]{lovelace76}
{Lovelace}, R.~V.~E. 1976, \nat, 262, 649

\bibitem[{{Marscher} {et~al.}(2008){Marscher}, {Jorstad}, {D'Arcangelo},
  {Smith}, {Williams}, {Larionov}, {Oh}, {Olmstead}, {Aller}, {Aller},
  {McHardy}, {L{\"a}hteenm{\"a}ki}, {Tornikoski}, {Valtaoja}, {Hagen-Thorn},
  {Kopatskaya}, {Gear}, {Tosti}, {Kurtanidze}, {Nikolashvili}, {Sigua},
  {Miller}, \& {Ryle}}]{mjd08}
{Marscher}, A.~P., {et~al.} 2008, \nat, 452, 966

\bibitem[{{Marscher} {et~al.}(2010){Marscher}, {Jorstad}, {Larionov}, {Aller},
  {Aller}, {L{\"a}hteenm{\"a}ki}, {Agudo}, {Smith}, {Gurwell}, {Hagen-Thorn},
  {Konstantinova}, {Larionova}, {Larionova}, {Melnichuk}, {Blinov},
  {Kopatskaya}, {Troitsky}, {Tornikoski}, {Hovatta}, {Schmidt}, {D'Arcangelo},
  {Bhattarai}, {Taylor}, {Olmstead}, {Manne-Nicholas}, {Roca-Sogorb},
  {G{\'o}mez}, {McHardy}, {Kurtanidze}, {Nikolashvili}, {Kimeridze}, \&
  {Sigua}}]{mjl10}
---. 2010, \apjl, 710, L126

\bibitem[{{McKinney} \& {Uzdensky}(2010)}]{mu10}
{McKinney}, J.~C., \& {Uzdensky}, D.~A. 2010, ArXiv e-prints

\bibitem[{{Murphy} {et~al.}(1991){Murphy}, {Cohn}, \& {Durisen}}]{mcd91}
{Murphy}, B.~W., {Cohn}, H.~N., \& {Durisen}, R.~H. 1991, \apj, 370, 60

\bibitem[{{Nakamura} {et~al.}(2006){Nakamura}, {McKee}, {Klein}, \&
  {Fisher}}]{nmk06}
{Nakamura}, F., {McKee}, C.~F., {Klein}, R.~I., \& {Fisher}, R.~T. 2006, \apjs,
  164, 477

\bibitem[{{O'Sullivan} \& {Gabuzda}(2009)}]{sg09}
{O'Sullivan}, S.~P., \& {Gabuzda}, D.~C. 2009, \mnras, 400, 26

\bibitem[{{Pittard} {et~al.}(2010){Pittard}, {Hartquist}, \& {Falle}}]{phf10}
{Pittard}, J.~M., {Hartquist}, T.~W., \& {Falle}, S.~A.~E.~G. 2010, \mnras,
  405, 821

\bibitem[{Ruffini \& Wilson(1975)}]{ruffini75}
Ruffini, R., \& Wilson, J.~R. 1975, Phys. Rev. D, 12, 2959

\bibitem[{{Savolainen} {et~al.}(2008){Savolainen}, {Wiik}, {Valtaoja}, \&
  {Tornikoski}}]{swvt08}
{Savolainen}, T., {Wiik}, K., {Valtaoja}, E., \& {Tornikoski}, M. 2008, in
  Astronomical Society of the Pacific Conference Series, Vol. 386,
  Extragalactic Jets: Theory and Observation from Radio to Gamma Ray, ed.
  {T.~A.~Rector \& D.~S.~De Young}, 451--+

\bibitem[{{Sikora}(2010)}]{s10psyn}
{Sikora}, M. 2010, ArXiv e-prints

\bibitem[{{Sokolovsky} {et~al.}(2010){Sokolovsky}, {Kovalev}, {Lobanov},
  {Savolainen}, {Pushkarev}, \& {Kadler}}]{skl10}
{Sokolovsky}, K.~V., {Kovalev}, Y.~Y., {Lobanov}, A.~P., {Savolainen}, T.,
  {Pushkarev}, A.~B., \& {Kadler}, M. 2010, ArXiv e-prints

\bibitem[{{Stern} \& {Poutanen}(2006)}]{sp06}
{Stern}, B.~E., \& {Poutanen}, J. 2006, \mnras, 372, 1217

\bibitem[{{Tavecchio} {et~al.}(2010){Tavecchio}, {Ghisellini}, {Bonnoli}, \&
  {Ghirlanda}}]{tgb10}
{Tavecchio}, F., {Ghisellini}, G., {Bonnoli}, G., \& {Ghirlanda}, G. 2010,
  \mnras, 405, L94

\bibitem[{{Tavecchio} {et~al.}(2009){Tavecchio}, {Ghisellini}, {Ghirlanda},
  {Costamante}, \& {Franceschini}}]{tgg09}
{Tavecchio}, F., {Ghisellini}, G., {Ghirlanda}, G., {Costamante}, L., \&
  {Franceschini}, A. 2009, \mnras, 399, L59

\bibitem[{{Tchekhovskoy} {et~al.}(2010){Tchekhovskoy}, {Narayan}, \&
  {McKinney}}]{tnm10}
{Tchekhovskoy}, A., {Narayan}, R., \& {McKinney}, J.~C. 2010, \na, 15, 749

\bibitem[{{Zacharopoulou} {et~al.}(2011){Zacharopoulou}, {Khangulyan},
  {Aharonian}, \& {Costamante}}]{zck10}
{Zacharopoulou}, O., {Khangulyan}, D., {Aharonian}, F.~A., \& {Costamante}, L.
  2011, \apj, 738, 157

\end{thebibliography}

\clearpage

\clearpage

\clearpage

\clearpage

\clearpage

\clearpage

\begin{figure}
\plotone{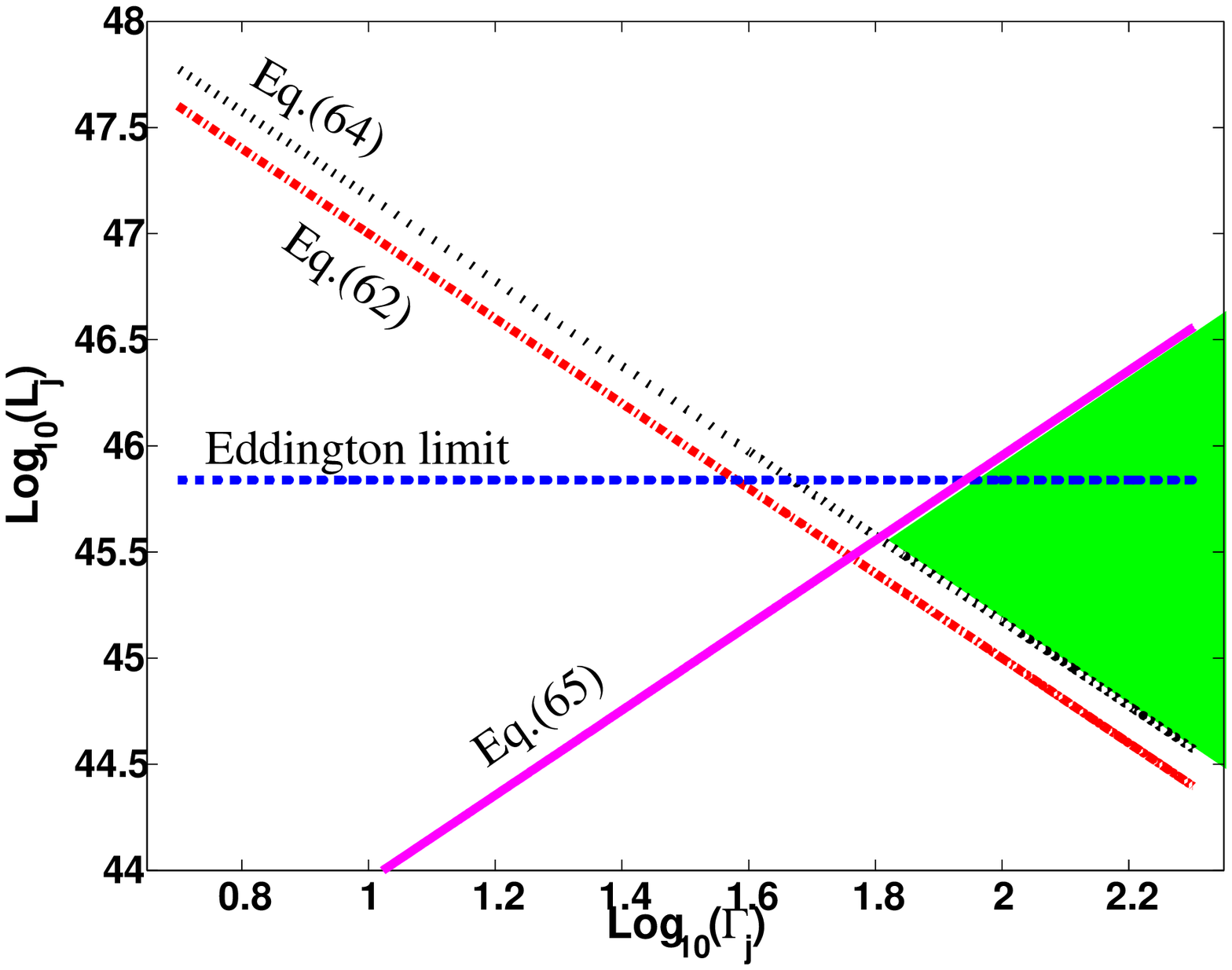}
\caption{Intensive VHE flares (in the case EIC) of 
$L_\gamma\approx10^{47}\,\rm erg\,s^{-1}$ and of duration $\tau\approx200$~s that
are energetically allowed in the framework of our JRGI scenario in systems 
characterized by a jet of luminosity, $L_{\rm j}$, propagating with bulk Lorentz factor $\Gamma_{\rm j}$,  are shown by the green filled region
(here we assume $M_{\rm BH}=5\times10^7M_\odot$ and $\xi_{-1}=1$).
The thin  solid line is the limit from Equation~(\ref{lj_condtion}); 
the thick  solid line is the limit from Equation~(\ref{lj_eic2}); 
the doted line is the limit from Equation~(\ref{lj_eic1}); 
the dot-dashed line is the limit from Equation~(\ref{jet_condtion}); and the dashed line 
is the Eddington luminosity. }
\label{lj_lor_eic}
\end{figure}
\clearpage

\clearpage

\begin{figure}
\plotone{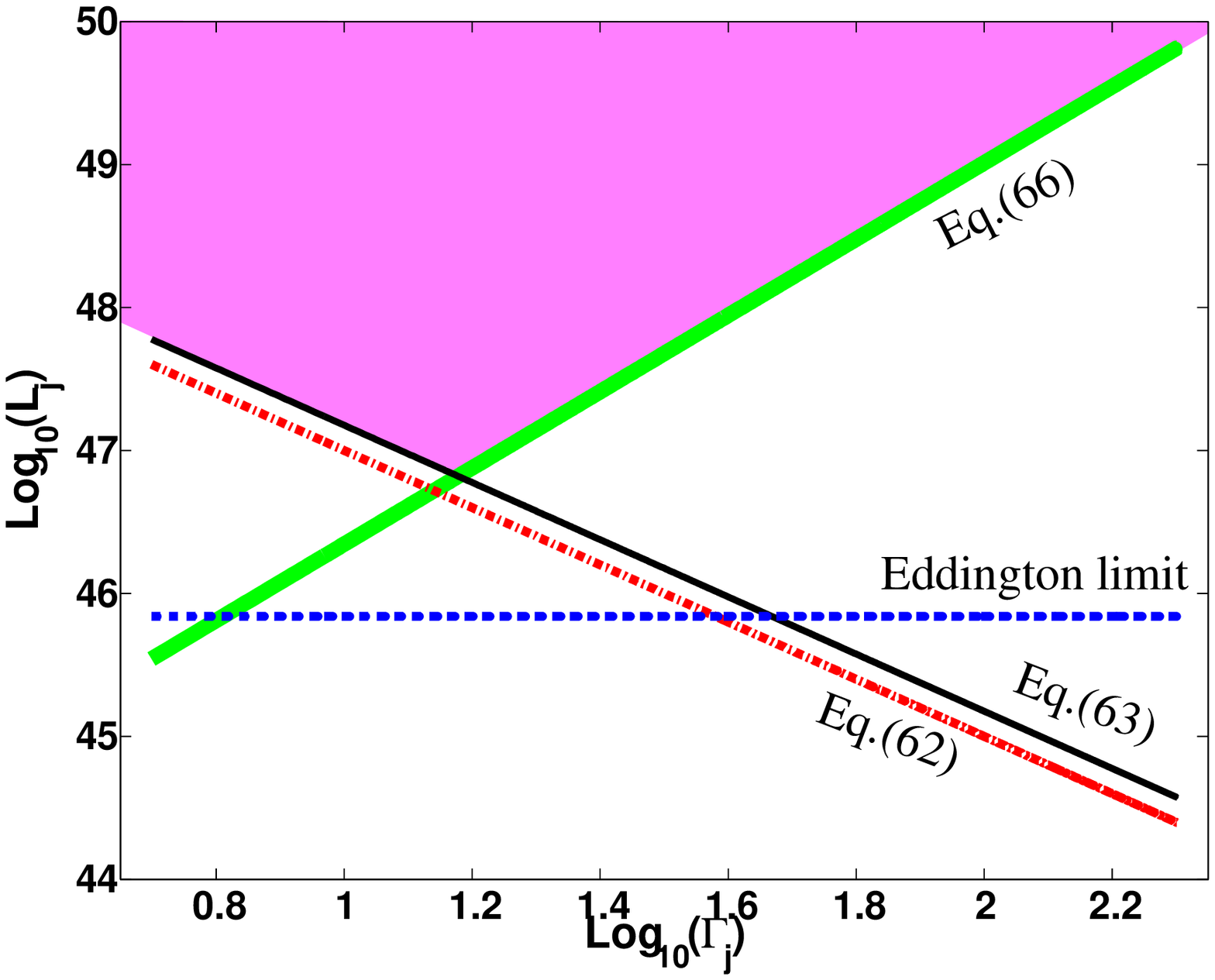}
\caption{Intensive VHE flares (in the case of proton synchrotron) 
of $L_\gamma\approx10^{47}\,\rm erg\,s^{-1}$ and of duration 
$\tau\approx200$~s that are energetically allowed in the framework 
of the JRGI scenario in systems characterized by a jet of luminosity, 
$L_{\rm j}$, propagating with bulk Lorentz factor $\Gamma_{\rm j}$, are shown by the pink filled region (we assume $M_{\rm BH}=5\times10^7M_\odot$ and $\xi_{-1}=1$).
The thin solid line is the limit from Equation~(\ref{lj_condtion}); 
the thick solid line is the limit from Equation~(\ref{lj_ps}); 
the dot-dashed line is the limit from Equation~(\ref{jet_condtion}); and the dashed line 
is the Eddington luminosity. }
\label{lj_lor_ps}
\end{figure}
\clearpage

\clearpage

\clearpage

\end{document}